\documentclass[aps,prb,twocolumn,superscriptaddress]{revtex4-2}
\usepackage{graphicx}
\usepackage{color,amssymb,amsmath}
\usepackage[normalem]{ulem}
\usepackage{hyperref}
\usepackage{comment}
\usepackage{lipsum}

\newcommand{\Deltait}{\mathit{\Delta}}
\newcommand{\Phio}{\mathit{\Phi}_0}

\newcommand{\PhiL}{\mathit{\Phi}_{\mathrm{L}}}
\newcommand{\PhiR}{\mathit{\Phi}_{\mathrm{R}}}
\newcommand{\IL}{I_{\mathrm{L}}}
\newcommand{\IR}{I_{\mathrm{R}}}
\newcommand{\Isd}{I_\mathrm{SD}}
\newcommand{\phiL}{\phi_\mathrm{L}}
\newcommand{\phiR}{\phi_\mathrm{R}}
\newcommand{\phiM}{\phi_\mathrm{M}}
\newcommand{\phiS}{\phi_\mathrm{S}}

\newcommand{\Isw}{I_{\mathrm{sw}}}
\newcommand{\Iswp}{I^+_{\mathrm{sw}}}
\newcommand{\Iswn}{I^-_{\mathrm{sw}}}
\newcommand{\Iswpn}{I^\pm_{\mathrm{sw}}}

\newcommand{\Icp}{I^+_{\mathrm{c}}}
\newcommand{\Icn}{I^-_{\mathrm{c}}}
\newcommand{\Icpn}{I^\pm_{\mathrm{c}}}

\newcommand{\Vj}{V_{\mathrm{J}}}
\newcommand{\Vl}{V_{\mathrm{L}}}
\newcommand{\Vr}{V_{\mathrm{R}}}
\newcommand{\Vs}{V_{\mathrm{S}}}
\newcommand{\Vm}{V_{\mathrm{M}}}

\begin{document}
\title{Flux-Tunable Josephson Diode Effect in a Hybrid Four-Terminal Josephson Junction}

\author{M.\ Coraiola}
\affiliation{IBM Research Europe---Zurich, 8803 R\"uschlikon, Switzerland}

\author{A.\ E.\ Svetogorov}
\affiliation{Fachbereich Physik, Universit\"at Konstanz, D-78457 Konstanz, Germany}

\author{D.\ Z.\ Haxell}
\affiliation{IBM Research Europe---Zurich, 8803 R\"uschlikon, Switzerland}

\author{D.\ Sabonis}
\affiliation{IBM Research Europe---Zurich, 8803 R\"uschlikon, Switzerland}

\author{M.\ Hinderling}
\affiliation{IBM Research Europe---Zurich, 8803 R\"uschlikon, Switzerland}

\author{S.\ C.\ ten Kate}
\affiliation{IBM Research Europe---Zurich, 8803 R\"uschlikon, Switzerland}

\author{E.\ Cheah}
\affiliation{Laboratory for Solid State Physics, ETH Z\"urich, 8093 Z\"urich, Switzerland}

\author{F.\ Krizek}
\altaffiliation[Present address: ]{Institute of Physics, Czech Academy of Sciences, 162 00 Prague, Czech Republic}
\affiliation{IBM Research Europe---Zurich, 8803 R\"uschlikon, Switzerland}
\affiliation{Laboratory for Solid State Physics, ETH Z\"urich, 8093 Z\"urich, Switzerland}

\author{R.\ Schott}
\affiliation{Laboratory for Solid State Physics, ETH Z\"urich, 8093 Z\"urich, Switzerland}

\author{W.\ Wegscheider}
\affiliation{Laboratory for Solid State Physics, ETH Z\"urich, 8093 Z\"urich, Switzerland}

\author{J.\ C.\ Cuevas}
\affiliation{Fachbereich Physik, Universit\"at Konstanz, D-78457 Konstanz, Germany}
\affiliation{Departamento de Física Te\'orica de la Materia Condensada and Condensed Matter Physics Center (IFIMAC), Universidad Aut\'onoma de Madrid, E-28049 Madrid, Spain}

\author{W.\ Belzig}
\affiliation{Fachbereich Physik, Universit\"at Konstanz, D-78457 Konstanz, Germany}

\author{F.\ Nichele}
\email{fni@zurich.ibm.com}
\affiliation{IBM Research Europe---Zurich, 8803 R\"uschlikon, Switzerland}

\date{\today}

\begin{abstract}
We investigate the direction-dependent switching current in a flux-tunable four-terminal Josephson junction defined in an InAs/Al two-dimensional heterostructure. The device exhibits the Josephson diode effect, with switching currents that depend on the sign of the bias current. The superconducting diode efficiency, reaching a maximum of $|\eta|\approx34\%$, is widely tunable---both in amplitude and sign---as a function of magnetic fluxes and gate voltages. Our observations are supported by a circuit model of three parallel Josephson junctions with non-sinusoidal current--phase relation. With respect to conventional Josephson interferometers, phase-tunable multiterminal Josephson junctions enable large diode efficiencies in structurally symmetric devices, where local magnetic fluxes break both time-reversal and spatial-inversion symmetries. Our work establishes a pathway to develop Josephson diodes with wide-range tunability and that do not rely on exotic materials or externally applied magnetic fields.
\end{abstract}

\maketitle

Nonreciprocal transport phenomena play a key role in modern electronics, with semiconductor diodes serving as the fundamental components for numerous devices \cite{Sze}. In analogy to the semiconductor diode, whose electrical resistance strongly depends on the current direction, a superconducting diode allows a larger supercurrent to flow in one direction compared to the other \cite{Nadeem2023}. Nonreciprocal supercurrents were recognized already in the 1970s in superconducting quantum interference devices (SQUIDs) based on superconducting bridges \cite{Fulton1970} and tunnel Josephson junctions (JJs) \cite{Fulton1972, Tsang1975}, arising as a consequence of the finite loop inductance. Direction-dependent switching currents were also observed in conventional superconducting thin films and interpreted as a manifestation of microscopic asymmetries in the device geometry \cite{Sivakov2018}. More recently, the superconducting diode effect (SDE) has sparked renewed interest, driven by its connection to fundamental properties of a diverse range of superconducting systems, where the breaking of both inversion and time-reversal symmetries is required for the effect to occur. Since its observation in artificial superlattices \cite{Ando2020}, the SDE has been the subject of thorough experimental and theoretical investigation, both in junction-free thin films \cite{Lyu2021, Daido2022, Ilic2022, He2022, Bauriedl2022, Hou2023} and JJs based on semiconductors with spin--orbit coupling \cite{Baumgartner2022, Baumgartner2022b, Costa2023, Lotfizadeh2023}, finite-momentum superconductors \cite{Davydova2022, Pal2022,Yuan2022} or multilayered materials, realizing sizeable asymmetries even without external magnetic fields \cite{Wu2022,Narita2022,Jeon2022,Lin2022,Zhang2022,Kokkeler2022,DiezMerida2023,Yun2023}.

An alternative platform proposed to achieve the SDE in Josephson devices---where it is usually referred to as Josephson diode effect (JDE)---relies on a supercurrent interferometer, where two JJs with nonsinusoidal current--phase relations (CPRs) are combined in a SQUID \cite{Souto2022, Fominov2022}. Such CPRs, containing contributions from higher harmonics than the conventional $2\pi$-periodic component, are routinely attained in high-quality superconductor--semiconductor planar materials \cite{English2016,Nichele2020,Aggarwal2021,Haxell2023b}, where hybrid JJs host Andreev bound states (ABSs) characterized by high transmission. Key ingredients for the JDE to occur in this system are the different harmonic content between the two JJs, and a magnetic flux threading the SQUID loop \cite{Souto2022}, as recently demonstrated in two-dimensional (2D) electron \cite{Ciaccia2023,Ciaccia2023b} and hole \cite{Valentini2023} systems, obtaining large diode efficiencies at equilibrium up to approximately $30\%$. Furthermore, three-terminal Josephson devices without phase control were shown to realize the JDE by exploiting high harmonic terms in the CPR \cite{Gupta2023}, multiple bias currents \cite{Chiles2023,Zhang2023b} and Andreev molecules \cite{Pillet2023,Matsuo2023b}.

In this work, we report on the superconducting transport properties of a superconductor--semiconductor four-terminal JJ (4TJJ) embedded in a double-loop geometry. Supercurrents are tuned by magnetic fluxes threading the superconducting loops and by gate electrodes that control the number and transmissions of ABSs. We find strong JDE, tunable in both efficiency and sign with gate voltages and two independent superconducting phases---controlled via integrated flux-bias lines, without the need for external magnetic fields---reaching a peak efficiency of $\pm 34\%$. One of the loops was further controlled by gating an additional JJ with large critical current, enabling operation of the device in a single-loop configuration. We provide an in-depth explanation of the JDE in our system by means of a simple circuit model, which maps our device to the combination of two SQUIDs, or a bi-SQUID. Simulations are performed both in an idealized case with minimal assumptions, and in an extended version that accurately captures the experimental results. We demonstrate several novelties with respect to previous work. First, we devise a geometry that combines both gate control of the current path and phase tunability in a 2D space, offering new insights into multiterminal JJs. Second, the phase-tunable bi-SQUID allows for a large diode efficiency to occur without the need of a JJ imbalance in harmonic content, which is required in conventional interferometers \cite{Souto2022}. Third, phase control enables large tuning of the JDE in amplitude and sign, including a vanishing diode efficiency in extended regions of the phase space. In the light of our results, multiterminal JJs in superconductor--semiconductor hybrid systems offer advantages that are pivotal for the realization of nonreciprocal transport phenomena. Thanks to the nonsinusoidal CPR and the possibility to break time-reversal symmetry solely by flux biasing, large and tunable diode efficiencies are naturally obtained without the need of an external magnetic field. Future work might further expand the study of multiterminal devices to realize nonreciprocal transport in the linear regime \cite{Leroux2022,Virtanen2023}, presenting opportunities for innovative applications.

\begin{figure}
	\includegraphics[width=\columnwidth]{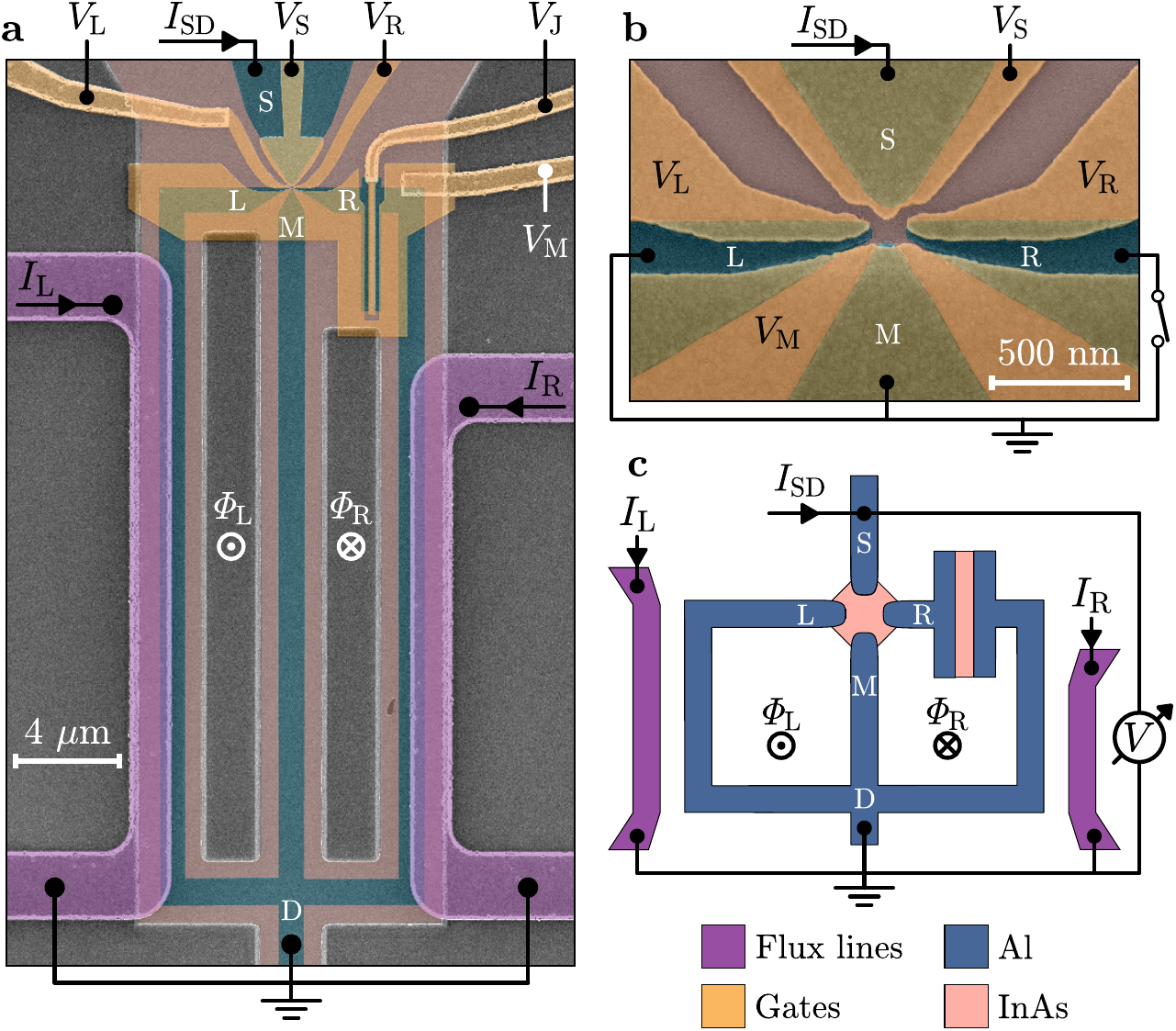}
	\caption{Device under study and measurement setup. (a) False-colored scanning electron micrograph of a device identical to that under study. Exposed III--V semiconductor is represented in pink, Al in blue, gate electrodes in yellow and flux-bias lines in purple. Bias current $\Isd$, flux-line currents $\IL$ and $\IR$, magnetic fluxes threading the superconducting loops $\PhiL$ and $\PhiR$, and gate voltages $\Vs$, $\Vl$, $\Vm$, $\Vr$ and $\Vj$ are labeled. Superconducting terminals S, L, M, R and the common node D are also indicated. (b) Zoom-in of (a) in the vicinity of the four-terminal Josephson junction. (c) Schematic representation of the device with the measurement setup, using the same color labeling as in (a) and (b). Gate electrodes are not shown.}
	\label{fig1}
\end{figure}

\begin{figure*}[t!]
	\includegraphics[width=\textwidth]{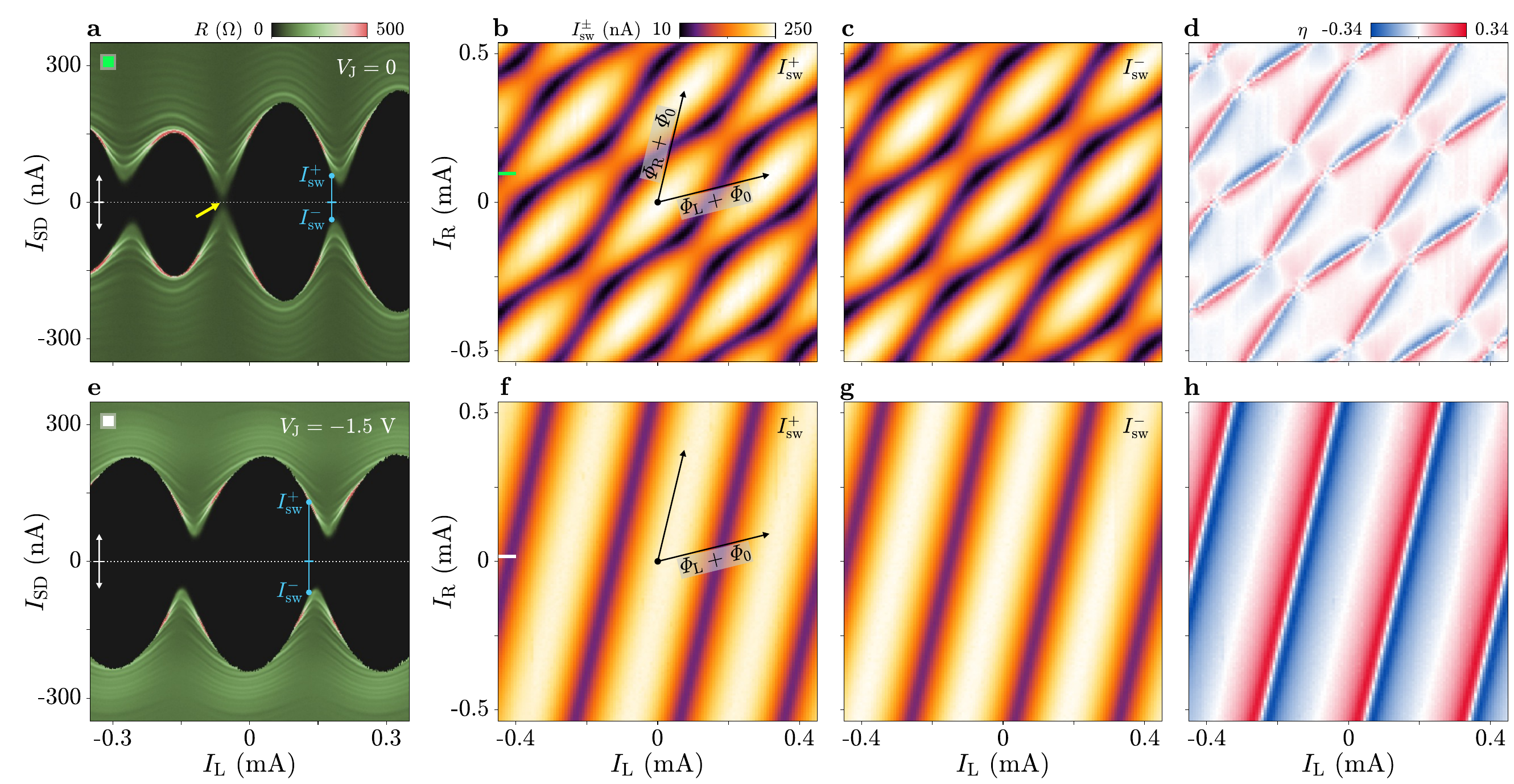}
	\caption{Phase-tunable Josephson diode effect. (a) Differential resistance $R$ as a function of left flux-line current $\IL$ and source--drain bias current $\Isd$, for fixed right flux-line current $\IR=0.1~\mathrm{mA}$. The map is obtained by merging two datasets recorded with $\Isd$ ramping from $0$ to either positive or negative values (see white arrows). Switching current nonreciprocities are highlighted by cyan annotations. A point where the switching current reaches zero is indicated by the yellow arrow. Gate voltages were set to $\Vl=\Vr=-0.1~\mathrm{V}$, $\Vs=0.1~\mathrm{V}$, $\Vm=-0.15~\mathrm{V}$ and $\Vj=0$. (b, c) Switching currents $\Iswp$ and $\Iswn$, measured for positive and negative $\Isd$ respectively, as functions of $\IL$ and $\IR$. Arrows in (b) indicate the directions along which magnetic fluxes threading the left and right superconducting loop, $\PhiL$ and $\PhiR$, vary independently. Each arrow represents the addition of one superconducting flux quantum $\Phio$ to the corresponding flux. (d) Superconducting diode efficiency $\eta$ calculated from (b) and (c) using Eq.~\ref{eq:eta} (see text), as a function of $\IL$ and $\IR$. (e--h) As in (a--d), but measured for $\Vj=-1.5~\mathrm{V}$, which sets the switch JJ to the OFF state and interrupts the right loop.}
	\label{fig2}
\end{figure*}

\section*{Results and Discussion}

\subsection*{Flux-tunable multiterminal Josephson junction}
The device under study, consisting of a multiterminal JJ embedded in a double-loop geometry, is displayed in Fig.~\ref{fig1}. It was realized in an InAs/Al heterostructure \cite{Shabani2016, Cheah2023}, where the epitaxial Al layer was selectively removed to expose the III--V semiconductor below. We defined four superconducting terminals, labeled S, L, M and R, coupled to a common semiconducting region. Lithographically, the minimum distances between neighboring terminals were $30~\mathrm{nm}$ (for L--M and R--M) and $50~\mathrm{nm}$ (for S--L and S--R), while opposite terminals had separations of $100~\mathrm{nm}$ (L--R) and $120~\mathrm{nm}$ (S--M). All junctions were short with respect to the superconducting coherence length in the InAs 2D electron gas, estimated to be approximately $600~\mathrm{nm}$ (see Methods section).
Terminals L, M and R were connected to a common node (D) forming two superconducting loops, which enabled independent control over two phase differences \cite{Coraiola2023, Coraiola2023b}, $\phiL-\phiM \equiv \phiL$ and $\phiR-\phiM \equiv \phiR$ (here, $\phi_\alpha$ indicates the superconducting phase of terminal $\alpha\in\{ \mathrm{L,M,R}\}$ and $\phiM$ was set to zero by convention). This was achieved by passing currents $\IL$ and $\IR$ through two flux-bias lines, patterned on top of a uniform dielectric layer, resulting in external magnetic fluxes $\PhiL$ and $\PhiR$ respectively threading the left and right loop. Gate electrodes were deposited on the same dielectric layer and energized by voltages $V_\alpha$ ($\alpha\in\{ \mathrm{S,L,M,R}\}$) and $\Vj$, allowing for electrostatic tuning of the electron density in the InAs layer below. While terminals L and M where directly connected to the node D via Al strips, a planar JJ (named switch JJ) was integrated on terminal R. The switch JJ, with a length of $40~\mathrm{nm}$ and a width of $5~\mathrm{\mu m}$, was designed to have a critical current much larger than that between any pairs of leads in the 4TJJ, and therefore the phase difference across the switch JJ can be neglected for the following discussion. Depending on the gate voltage $V_\mathrm{J}$, the switch JJ was employed in two configurations: $\Vj=0$ (switch ON), where the JJ was conducting and $\PhiR$ could be used to control $\phiR$, or $\Vj=-1.5~\mathrm{V}$ (switch OFF), where the JJ was depleted, the right loop was interrupted and terminal R was reduced to a floating superconducting island. The other gate voltages were set to $\Vs=0.1~\mathrm{V}$, $\Vl=\Vr=-0.1~\mathrm{V}$ and $\Vm=-0.15~\mathrm{V}$, unless stated otherwise. The device was measured in a dilution refrigerator with a base temperature of about $10 ~\mathrm{mK}$. Current-bias experiments were performed in a four-terminal configuration by driving a current $\Isd$ between S and D and measuring the voltage drop across the device, which allowed for the measurement of the switching current $\Isw$. Along its path between S and D, the current flowed through the semiconducting region forming the four-terminal JJ, and in particular across the S--L, S--M and S--R junctions. 
In our geometry, the superconducting loops were designed to limit their inductance, that could in principle lead to the SDE in the system. The maximum flux variation due to the inductance of a loop, estimated to be approximately $124~\mathrm{pH}$ (see details in the Supplementary Material, Section 5), for a circulating current of the order of $100~\mathrm{nA}$, is $\sim 6 \times 10^{-3} \Phio$. This was observed to be negligible with respect to the flux scales over which the device properties varied.
Further details regarding materials, fabrication and measurement setup are provided in the Methods section. Results on a second device, similar to the one discussed in the Main Text, are presented in the Supplementary Material (see Figs.~S6--S10 in Section 4). Devices studied here were employed in a previous work that investigated hybridization of ABSs in multiterminal JJs \cite{Coraiola2023}.

\subsection*{Nonreciprocal supercurrents in the 2D phase space}
First, we present the differential resistance $R$ as a function of the current bias $\Isd$ and of the left flux-line current $\IL$ for fixed right flux-line current $\IR=0.1~\mathrm{mA}$. Here, $R$ was measured with standard lock-in techniques and $\Isd$ was swept from $0$ to positive or negative values to avoid retrapping effects.
Figure \ref{fig2}a shows the result for $\Vj=0$ (switch ON): the switching current revealed oscillations of varying amplitude as a function of $\IL$, which, notably, were nonreciprocal at positive and negative $\Isd$. For instance, at $\IL=0.18~\mathrm{mA}$ we measured switching currents $\Iswp=58~\mathrm{nA}$ at $\Isd>0$ and $\Iswn=38~\mathrm{nA}$ at $\Isd<0$ (see cyan annotations), where $\Iswpn>0$ by definition. This resulted in a superconducting diode efficiency $\eta$, defined as:
\begin{equation}\label{eq:eta}
	\eta = \frac{\Iswp - \Iswn}{\Iswp + \Iswn},
\end{equation}
of approximately $21 \%$. We also note that the switching current vanished in a small range around $\IL \approx -60~\mathrm{\mu A}$ (yellow arrow), namely the device had finite differential resistance at $\Isd=0$. Similar maps obtained at different settings of $\IR$ are presented in Fig.~S1 of the Supplementary Material.

To efficiently measure the switching currents $\Iswpn$ and the diode efficiency $\eta$ as functions of both $\IL$ and $\IR$, we changed measurement technique and periodically ramped $\Isd$ from zero to the amplitude $A = \pm 260~\mathrm{nA}$ with a repetition rate of $133~\mathrm{Hz}$, and detected when the voltage drop across the device exceeded a threshold. The time spent in the low-resistance state, averaged over 32 consecutive measurements, was converted to a current, resulting in a rapid measurement of $\Iswp$ or $\Iswn$ (depending on the sign of $A$) displayed in Figs.~\ref{fig2}b and \ref{fig2}c, respectively. A limitation of this measurement technique was that values of $\Isw$  below $\sim 10~\mathrm{nA}$ could not be accurately detected due to the finite voltage threshold, which gave a finite reading of about $10~\mathrm{nA}$ for small switching currents, and even when the device was resistive for zero bias current.
The switching current oscillated periodically in the 2D phase space---where the periodicity axes correspond to the external magnetic fluxes $\PhiL$ and $\PhiR$ (black lines in Fig.~\ref{fig2}b)---forming a pattern characterized by lobe-like features. The finite slope of the $\PhiL$ and $\PhiR$ axes with respect to $\IL$ and $\IR$ was due to the cross-coupling between the left (right) flux-bias line and the right (left) loop, as discussed in Section 5 of the Supplementary Material. The oscillations of $\Isw$ exhibited maxima of approximately $250~\mathrm{nA}$ where $\PhiL$ and $\PhiR$ were integer multiples of the superconducting flux quantum $\Phio=h/2e$ (with $h$ the Planck constant and $e$ the elementary charge), and minima at finite phases where the limit of detection was reached, consistent with the vanishing switching current discussed for Fig.~\ref{fig2}a. We note that the switching currents were nonreciprocal upon reversal of the current bias, while their 2D patterns were symmetric to each other with respect to the origin ($\IL=\IR=0$, corresponding to $\PhiL=\PhiR=0$). The symmetry was particularly visible in the shape of the lobes, which was inverted as the supercurrent changed sign.
Figure \ref{fig2}d shows the superconducting diode efficiency calculated from Figs.~\ref{fig2}b and \ref{fig2}c using Eq.~\ref{eq:eta}. As expected, $\eta$ reflected the 2D periodic pattern in the phase space of the switching currents, and was widely tunable as a function of $\IL$ and $\IR$. We observe a fully ambipolar character and large values up to $\eta \approx \pm 21 \%$ where $\Iswpn$ had a large gradient in the phase space, while the efficiency vanished in extended regions of the phase space without the need for fine tuning $\IL$ and $\IR$.

\subsection*{Nonreciprocal supercurrents in single-loop configuration}
Next, in Figs.~\ref{fig2}e--h we present the measurements corresponding to those discussed in Figs.~\ref{fig2}a--d but with the switch junction in the OFF state ($\Vj=-1.5~\mathrm{V}$). From the differential resistance as a function of $\Isd$ and $\IL$ (Fig.~\ref{fig2}e, here for $\IR=17~\mathrm{\mu A}$), we found periodic oscillations of the switching current, with $\Iswp$ and $\Iswn$ exhibiting a phase shift from each other and opposite skewness (in the forward direction for $\Iswp$, backward for $\Iswn$). Consequently, the switching currents were again nonreciprocal depending on $\Isd$, which indicates a large JDE; at $\IL=0.13~\mathrm{mA}$, for example, $\Iswp=130~\mathrm{nA}$ and $\Iswn=68~\mathrm{nA}$ (see cyan annotations), yielding $\eta \approx 31\%$. We note that, in this configuration, the switching current did not vanish for any value of $\IL$, with minimal values of $\approx 50~\mathrm{nA}$, unlike the case with the switch ON.
Measurements of $\Iswp$ and $\Iswn$ as functions of both $\IL$ and $\IR$ are shown in Figs.~\ref{fig2}f and \ref{fig2}g. The 2D pattern observed for $\Vj=0$ was no longer present: as expected, the dependence on the flux $\PhiR$ was suppressed when the right superconducting loop was interrupted, and periodicity remained along a single direction. In agreement with Fig.~\ref{fig2}e, the switching current oscillations were shifted in phase and their skewness was reversed depending on the sign of the current bias. The superconducting diode efficiency, displayed in Fig.~\ref{fig2}h for the data of panels f and g, was also characterized by a periodic behavior as a function of $\PhiL$ and reached maxima of approximately $34\%$.

\begin{figure}[t!]
	\includegraphics[width=\columnwidth]{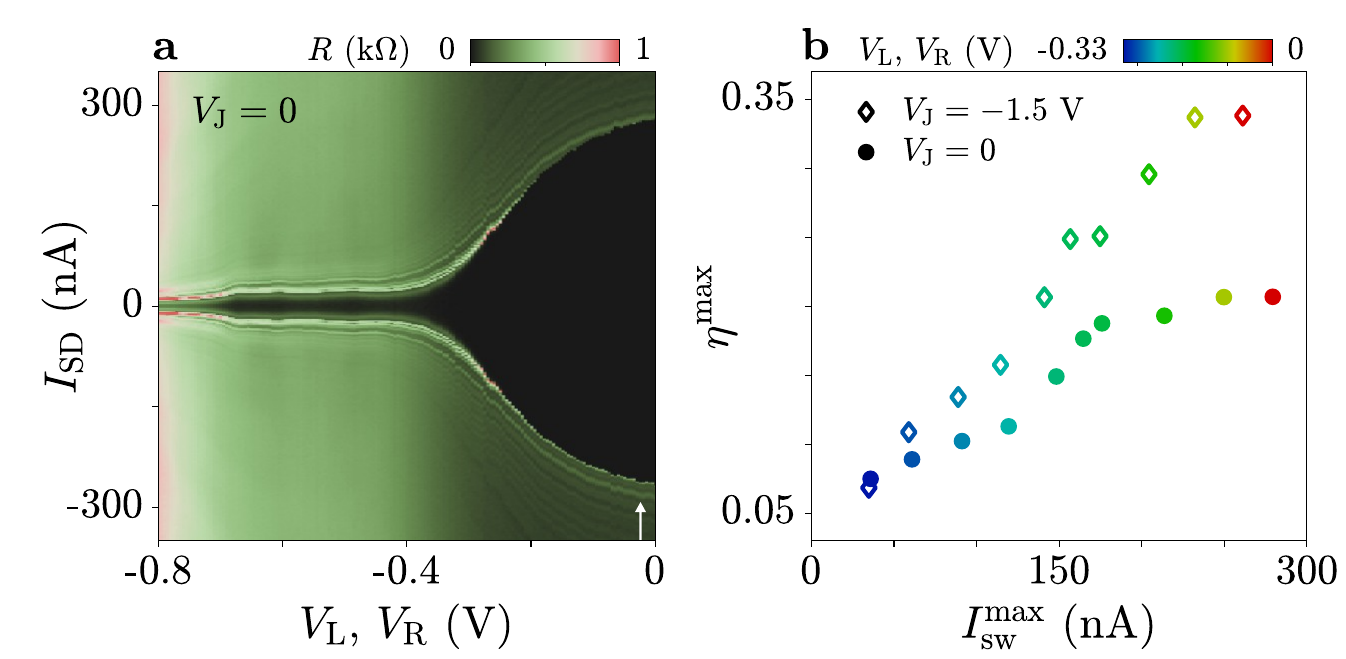}
	\caption{Gate-tuning of the maximum switching current and diode efficiency. (a) Differential resistance $R$ as a function of gate voltages $V_{\mathrm{L}}=V_{\mathrm{R}}$ and bias current $\Isd$ (swept from negative to positive values, see white arrow). (b) Maximum Josephson diode efficiency $\eta^{\mathrm{max}}$ as a function of the maximum switching current $I_\mathrm{sw}^\mathrm{max}$ for multiple values of $V_{\mathrm{L}}=V_{\mathrm{R}}$ (see colorscale). Each point is obtained from maps similar to Figs.~\ref{fig2}b and \ref{fig2}c (see Supplementary Material, Section 3 for more details). Full circles refer to the situation with $\Vj=0$, empty diamonds to the situation with $\Vj=-1.5~\mathrm{V}$.
	}
	\label{fig3}
\end{figure}

\begin{figure*}[t!]
	\includegraphics[width=0.75\textwidth]{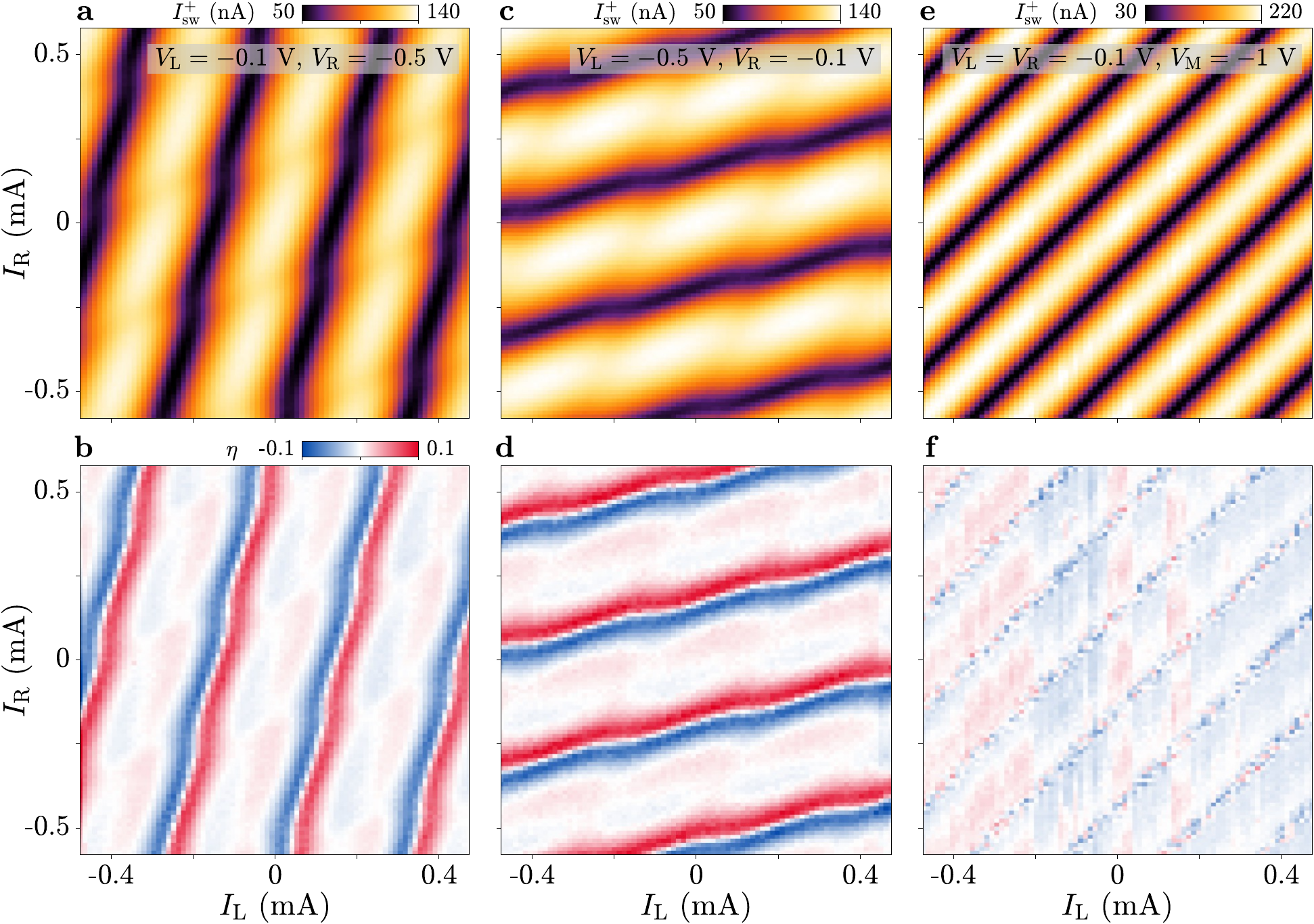}
	\caption{Routing of the supercurrent. (a) Switching current $\Iswp$ measured for positive bias current $\Isd$, as a function of flux-line currents $\IL$ and $\IR$. Measurements are performed with $\Vl=-0.1~\mathrm{V}$ and $\Vr=-0.5~\mathrm{V}$. (b) Diode efficiency $\eta$ for the configuration of (a). (c, d) As in (a, b), but for $\Vl=-0.5~\mathrm{V}$ and $\Vr=-0.1~\mathrm{V}$. (e, f) As in (a, b), but for $\Vl=-0.1~\mathrm{V}$, $\Vr=-0.1~\mathrm{V}$, and $\Vm=-1~\mathrm{V}$.
	}
	\label{fig4}
\end{figure*}

\subsection*{Gate-tunable Josephson diode effect}

Electrostatic tunability over the supercurrents and the JDE was enabled by gates controlling the electron density in the semiconducting region of the 4TJJ. In Fig.~\ref{fig3}a, we show the differential resistance as a function of the current bias (here swept from negative to positive values, thus displaying both retrapping and switching currents) while the gate voltages $\Vl$ and $\Vr$ varied simultaneously, for $\Vj=0$ and $\IL=\IR=0$. The switching current and the retrapping current, respectively $280~\mathrm{nA}$ and $-265~\mathrm{nA}$ at $\Vl=\Vr=0$, decreased for more negative voltages, until a finite resistance was measured at $\Isd=0$ for $\Vl=\Vr\approx-0.35~\mathrm{V}$. To investigate the voltage-tunability of the JDE, we measured the switching currents $\Iswpn$ as functions of $\IL$ and $\IR$ (as in Figs.~\ref{fig2}b, c, f, g) for varying $\Vl$ and $\Vr$, and, in each configuration, we extracted the peak values of $\Iswp$, named $I_\mathrm{sw}^\mathrm{max}$, and of the superconducting diode efficiency, $\eta^\mathrm{max}$ (see Supplementary Material, Section 3 for the details of the extraction procedure). The result is presented in Fig.~\ref{fig3}b, where $\eta^\mathrm{max}$ is plotted as a function of $I_\mathrm{sw}^\mathrm{max}$ for the two settings of the switch JJ $\Vj=0$ and $\Vj=-1.5~\mathrm{V}$. In both cases, we observed an increasing trend of $\eta^\mathrm{max}$ as $I_\mathrm{sw}^\mathrm{max}$ increased. For any gate setting, the diode efficiency was larger at $\Vj=-1.5~\mathrm{V}$ than at $\Vj=0~\mathrm{V}$, up to $34 \%$ and $21 \%$ respectively (at $\Vl=\Vr=0$), while $I_\mathrm{sw}^\mathrm{max}$ was slightly smaller in the former case (up to $260~\mathrm{nA}$, instead of $280~\mathrm{nA}$ for $\Vj=0$).

We further characterized the gate dependence of the device by allowing an asymmetric tuning of $\Vl$ and $\Vr$ (at $\Vj=0$), as shown in Figs.~\ref{fig4}a, b and \ref{fig4}c, d for the configurations $\Vl=-0.1~\mathrm{V},~\Vr=-0.5~\mathrm{V}$ and $\Vl=-0.5~\mathrm{V},~\Vr=-0.1~\mathrm{V}$. In each case, the switching current measured as a function of $\IL$ and $\IR$ for positive current bias is displayed in the first panel, while the second panel presents the diode efficiency extracted from $\Iswp$ and $\Iswn$. The two configurations revealed a complementary behavior: the switching current oscillations and the diode efficiency were almost completely suppressed as a function of $\mathit{\Phi}_\mathrm{L(R)}$ when $V_\mathrm{L(R)}$ was set to sufficiently negative value, depleting the semiconducting region between terminals S and L (R). This highlights the possibility of routing the supercurrents flowing in our device by gating, which enabled electrostatic control over the phase dependence of the JDE. The results obtained for $\Vr=-0.5~\mathrm{V}$ (panels a and b) were reminiscent of those previously observed for $\Vj=-1.5~\mathrm{V}$ (Figs.~\ref{fig2}f--h), where data were independent of $\PhiR$.

Finally, we restored the symmetric gate configuration $\Vl=\Vr=-0.1~\mathrm{V}$ and studied the effect of depleting the middle gate $\Vm$, set to $-1~\mathrm{V}$ (see Fig.~\ref{fig4}e and \ref{fig4}f). Here, we observed periodic oscillations of the switching current along a single direction of the phase space, corresponding to the $(\PhiL-\PhiR)$-axis. The frequency of these oscillations was doubled compared to the case where $\Vm$ was not depleted (e.g., Fig.~\ref{fig2}b), consistent with the exclusion of terminal M from the current path. As a consequence, screening currents induced by the flux-bias lines only circulated in the perimeter of the double-loop geometry, leading $\IL$ and $\IR$ to control the total flux $\PhiL-\PhiR$ (note that $\PhiL$ and $\PhiR$ were defined with opposite signs in Figs.~\ref{fig1}a and \ref{fig1}c). Notably, in this symmetric gate configuration where no current flowed into terminal M, the JDE was essentially suppressed (see Fig.~\ref{fig4}f).
Results for additional gate settings are shown in the Supplementary Material, see Figs.~S2--S5.

\begin{figure*}[t!]
	\includegraphics[width=\textwidth]{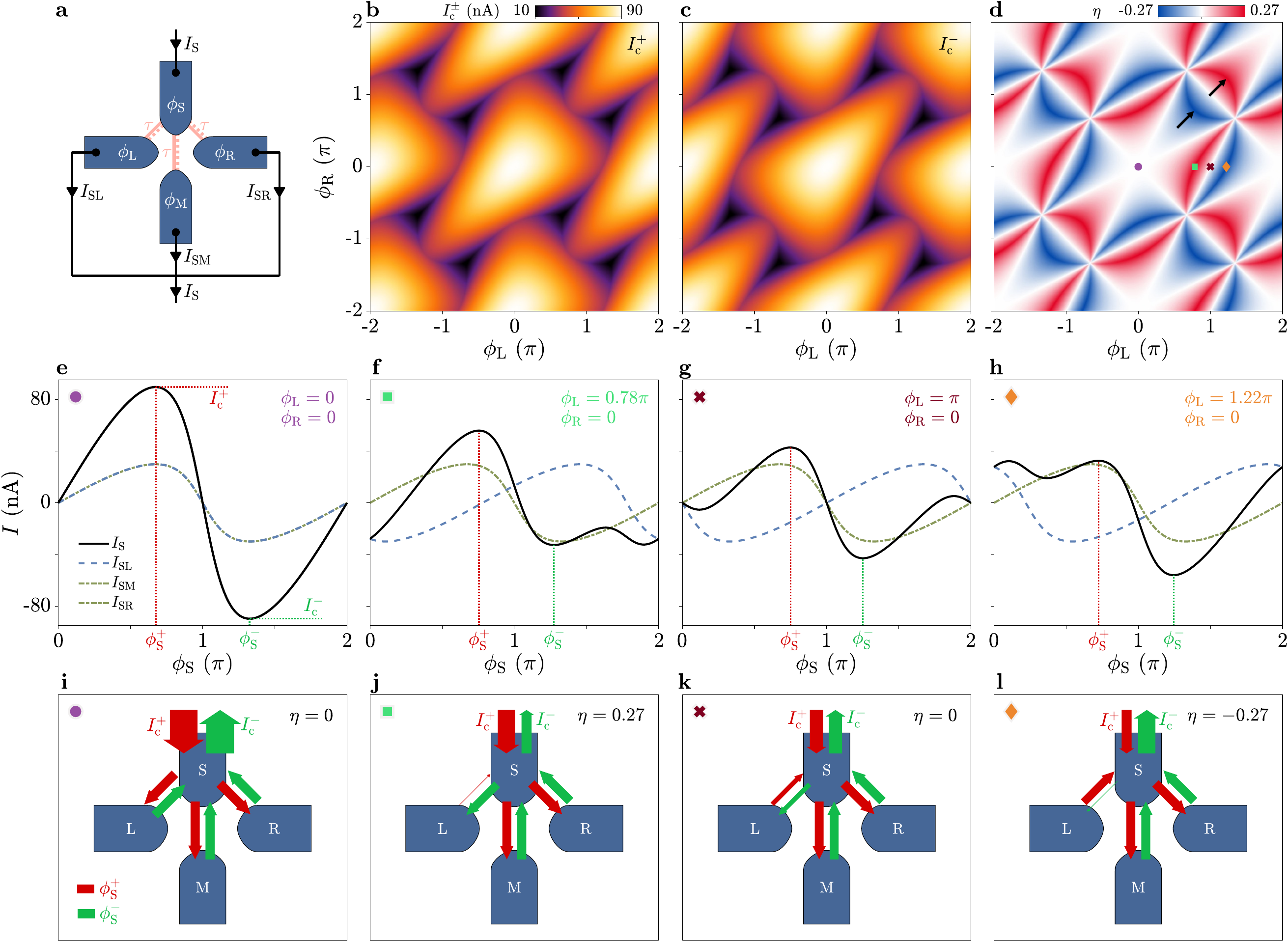}
	\caption{Minimal-model description of the Josephson diode effect. (a) Schematic representation of the four-terminal Josephson junction and circuit parameters. (b, c) Simulated critical currents $\Icp$ and $\Icn$ for positive and negative $\Isd$, respectively, as functions of the phase differences $\phiL$ and $\phiR$. The transmission of the three channels is $\tau=0.9$. (d) Diode efficiency $\eta$ derived from (b) and (c) as a function of $\phiL$ and $\phiR$. (e--h) Supercurrents in the four leads as functions of phase $\phiS$. The phases of the other leads are indicated in each panel. The four cases correspond to the colored markers in (d). The value of $\phiS$ where $I_\mathrm{S}$ has its maximum $\Icp$ (minimum $\Icn$), labeled $\phiS^+$ ($\phiS^-$), is highlighted by the red (green) dotted line. (i--l) Schematic representation of the supercurrent flow in the phase configurations shown in (e--h). Red and green arrows indicate supercurrents for $\phiS^+$ and $\phiS^-$, respectively. The wider the arrow, the larger the supercurrent.}
	\label{fig5}
\end{figure*}

\subsection*{Minimal-model description of JDE}
To understand the behavior of our device in more depth, and the underlying origin of the JDE, we introduce a simple circuit model that describes the supercurrents of the 4TJJ. In the switch-ON configuration, terminals L, M and R are all connected to the same node D, therefore the current can flow from S to D (or from D to S) via three available JJs S--L, S--M and S--R. That is, the 4TJJ is mapped onto a bi-SQUID as three distinct JJs are connected in parallel. The total current flowing into lead S is thus expressed as:
\begin{equation}\label{eq:Itot}
	I_\mathrm{S} = I_\mathrm{SL} + I_\mathrm{SM} + I_\mathrm{SR} = \sum_{\alpha} I_\mathrm{S \alpha},
\end{equation}
where $I_\mathrm{S \alpha}$ is the current flowing from terminal S to terminal $\alpha$ via the corresponding JJ.
First, we consider the minimal model of a single numerical parameter, schematically shown in Fig.~\ref{fig5}a. Each of the three JJs, that are identical to each other, is described by one conductive channel with transmission $\tau$, resulting in the CPR \cite{Beenakker1991b}:
\begin{equation}\label{eq:CPR:1}
	I_\mathrm{S \alpha} = \frac{e \Deltait}{2 \hbar} \frac{\tau \sin (\phiS-\phi_\alpha)}{\sqrt{1-\tau \sin^2 \frac{\phiS-\phi_\alpha}{2} }}.
\end{equation}
with $\Deltait$ the superconducting gap ($\Deltait = 180~\mathrm{\mu eV}$ is used considering Al leads) and $\hbar=h/2\pi$.
We assume a high transmission $\tau = 0.9$, for which the CPR of Eq.~\ref{eq:CPR:1} has a significantly nonsinusoidal character, as harmonics higher than the first provide a sizable contribution. The independent variables in the model are the three superconducting phase differences $\phiL$, $\phiR$ and $\phiS$, defined with respect to $\phiM \equiv 0$. The first two phases are related to the external magnetic fluxes by $\phi_\mathrm{L(R)}=2 \pi  \mathit{\Phi}_\mathrm{L(R)} / \Phio$ (neglecting the inductance of the loops, see discussion in the Supplementary Material, Section 5), whereas $\phiS$ varies depending on current bias $\Isd$. The critical currents for the two bias directions are then obtained as:
\begin{equation}\label{eq:Icpn}
	I_\mathrm{c}^\pm = \max_{\phi_\mathrm{S}} \left[ \pm I_\mathrm{S}(\phi_\mathrm{S}) \right].
\end{equation}
In Figs.~\ref{fig5}b and \ref{fig5}c, we show $\Icpn$ computed as functions of $\phiL$ and $\phiR$, while the diode efficiency calculated with Eq.~\ref{eq:eta} (where $\Iswpn$ are substituted by $\Icpn$) is displayed in Fig.~\ref{fig5}d. The critical currents, fulfilling the condition $\Icp(\phiL,\phiR)=\Icn(-\phiL,-\phiR)$, exhibit patterns that are prominently asymmetric with respect to $\phiL=\phiR=0$ (modulo $2 \pi$), which leads to a strong JDE with $\eta$ up to approximately $27\%$. The dependence of $\eta$ on $\phiL$ and $\phiR$ reflects the triangular shapes observed for $\Icpn$, with features arranged according to three main orientations in the phase space. 
The origin of the JDE is investigated by fixing $\phiL$ and $\phiR$ and computing the CPR of Eq.~\ref{eq:Itot} as a function of $\phiS$, $I_\mathrm{S}(\phiS)$, and its components $I_\mathrm{SL}(\phiS)$, $I_\mathrm{SM}(\phiS)$ and $I_\mathrm{SR}(\phiS)$, all obtained from Eq.~\ref{eq:CPR:1}. For simplicity we always keep $\phiR=0$ and select four values of $\phiL$ (colored markers in Fig.~\ref{fig5}d), where $|\eta|$ is either zero ($\phiL=0,\pi$) or maximal ($\phiL=0.78\pi, 1.22 \pi$). The individual and combined CPRs at these phase-space points are plotted in Fig.~\ref{fig5}e--h. In each case, we identify the values of $\phiS$ that maximize the total current $I_\mathrm{S}$ ($\phiS^+$, red dotted lines) and its inverse $-I_\mathrm{S}$ ($\phiS^-$, green dotted lines), such that $I_\mathrm{S}(\phiS^\pm)=\Icpn$. The currents flowing to and from terminal S are schematically depicted in Fig.~\ref{fig5}i--l for the same $\phiL$ and $\phiR$ values of panels e--h. In the schematics, red (green) arrows show the situation at $\phiS^{+(-)}$, their width and direction indicate the magnitude and direction of the current.
We note that all individual CPRs $I_\mathrm{S \alpha}(\phiS)$ have the same amplitude $\approx 30~\mathrm{nA}$ and skewness, both given by the transmission $\tau$ (identical for the three channels), but notably $I_\mathrm{S L}(\phiS)$ is phase-shifted by $\phiL$. When $\phiL=0$ (Fig.~\ref{fig5}e, i), all components are in-phase and $I_\mathrm{S}(\phiS) = 3 I_\mathrm{S \alpha}(\phiS)$, hence a standard nonsinusoidal CPR is obtained. Positive and negative critical currents are identical, and all currents are simply reversed between $\phiS^+$ and $\phiS^-$. In contrast, when $I_\mathrm{SL}(\phiS)$ is shifted by $\phiL=0.78 \pi$ (Fig.~\ref{fig5}f, j), the total CPR becomes nonreciprocal for positive and negative currents. The $I_\mathrm{SL}$-component is very small at $\phiS^+$, but comparable with $I_\mathrm{SM,SR}$ at $\phiS^-$; since in both cases $I_\mathrm{SL}$ has opposite sign with respect to $I_\mathrm{SM,SR}$, this asymmetry leads to $\Icp > \Icn$.
A symmetric scenario is recovered for $\phiL=\pi$, when $I_\mathrm{SL}$ is shifted by half period from the other components. Here, $I_\mathrm{S}$ always has the opposite sign to $I_\mathrm{SM,SR}$, but same absolute value at $\phiS^\pm$, such that $\Icp = \Icn$ and no JDE is present. Finally, the results obtained for $\phiL=1.22 \pi$ (Fig.~\ref{fig5}h,l), symmetric about $\phiL=\pi$ to $\phiL=0.78 \pi$, show the same CPRs discussed for Fig.~\ref{fig5}f,j upon sign inversion of both current and $\phiS$, confirming that here the JDE is strong and has opposite direction compared to the previous case.

We note that the JDE has been derived within our minimal model despite the presence of three identical channels, i.e., of same harmonic content, whereas in a conventional SQUID comprising two JJs the harmonic content must be different between the two JJs \cite{Souto2022}. The multiterminal geometry that we discuss can also be reduced to a conventional SQUID, where two JJs and a phase degree of freedom (for example, the S--M and S--R junctions and $\phiR$) are replaced by an effective JJ of tunable harmonic content. The effective JJ, together with the third JJ and the remaining phase difference (in the same example, S--L and $\phiL$), realizes the proposal of Ref.~\cite{Souto2022}. An important advantage offered by our platform lies in the possibility of phase-tuning the harmonic content of the effective junction, establishing wide and flexible control over the JDE. The nonsinusoidal character of the individual CPRs, which is still a requirement, is obtained in high-transmission hybrid JJs (as those realized in this work), while our novel geometry eliminates the need for precise control over the transmissions of the single junctions after device fabrication.

\subsection*{Simulations with the extended model}

\begin{figure*}[t!]
	\includegraphics[width=\textwidth]{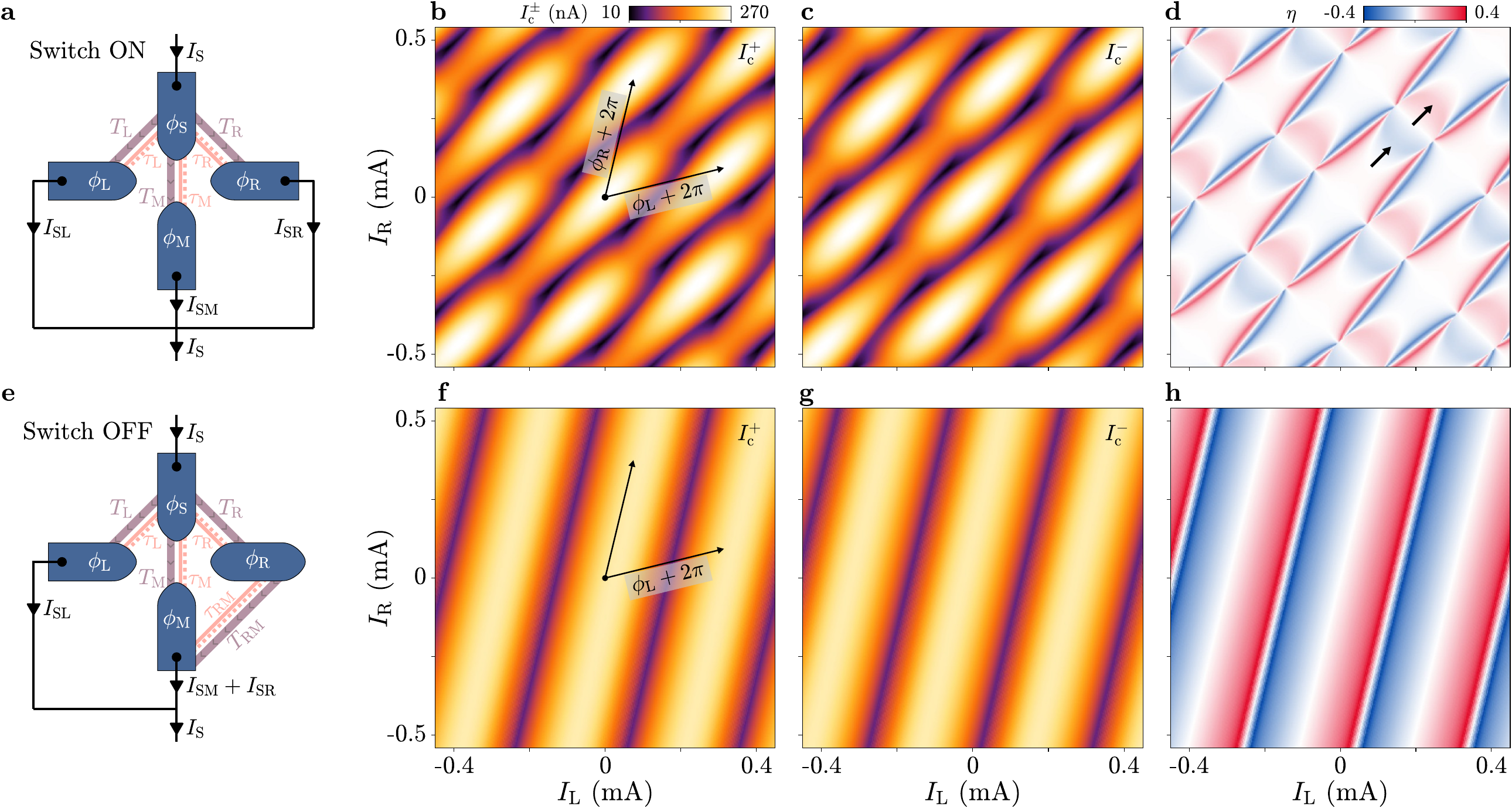}
	\caption{Josephson diode effect in the extended model. (a) Schematic representation of the four-terminal Josephson junction and circuit parameters for the switch-ON configuration (see text for details). (b, c) Simulated critical currents $\Icp$ and $\Icn$ for positive and negative current bias (respectively), as functions of the flux-line currents $\IL$ and $\IR$. Currents $\IL$ and $\IR$ are calculated from the superconducting phase differences $\phiL$, $\phiR$ and the mutual  inductance matrix (see discussion in the Supplementary Material, Section 5). Black arrows, whose directions indicate the periodicity axes $\phiL$ and $\phiR$, represent the corresponding phase winding by $2 \pi$. (d) Diode efficiency $\eta$ obtained from (b) and (c) as a function of $\IL$ and $\IR$. (e) Schematic representation of the four-terminal Josephson junction and circuit parameters for the switch-OFF configuration. (f, g) As in (b, c), but for the case with switch-OFF. (h) Diode efficiency $\eta$ obtained from (f) and (g) as a function of $\IL$ and $\IR$.}
	\label{fig6}
\end{figure*}

After discussing the origin of the JDE by means of a minimal, single-parameter model, we extend the latter to better capture the experimental data presented in Fig.~\ref{fig2}. Again, we consider the three JJs S--L, S--M and S--R that contribute to the total current according to Eq.~\ref{eq:Itot}. For each junction we consider two contributions to the current $I_\mathrm{S \alpha}$: in addition to a high-transparency channel, with transmission $\tau_{\alpha}$, a component with conventional sinusoidal CPR \cite{Tinkham2004}, associated to a large number of low-transmission channels, is included:
\begin{equation}\label{eq:CPR:2}
	I_\mathrm{S \alpha} = \frac{e \Deltait}{2 \hbar} \left[ \frac{\tau_\alpha \sin (\phi_\mathrm{S}-\phi_\alpha)}{\sqrt{1-\tau_\alpha \sin^2 \frac{\phi_\mathrm{S}-\phi_\alpha}{2} }} + T_\alpha \sin (\phi_\mathrm{S}-\phi_\alpha)\right],
\end{equation}
where $e \Deltait T_\alpha / 2 \hbar$ is the critical current of the sinusoidal component and $T_\alpha$ the sum of the transmissions of all low-transmission channels. We note that $\tau_{\alpha}$ and $T_{\alpha}$ may vary depending on the junction S--$\alpha$. With this extended model, that is schematically represented in Fig.~\ref{fig6}a, we compute the critical currents using Eq.~\ref{eq:Icpn} for any settings of $\phiL$ and $\phiR$. The simulated $\Icp$, $\Icn$ and $\eta$ are shown in Fig.~\ref{fig6}b--d for parameters $\tau_\mathrm{L}=\tau_\mathrm{R}=0.92$, $\tau_\mathrm{M}=0.89$, $T_\mathrm{L}=3.5$, $T_\mathrm{M}=1.5$ and $T_\mathrm{R}=3.6$. For a better comparison to the experimental results, the three quantities are plotted as functions of the flux-bias line currents $\IL$ and $\IR$, calculated from the phases $\phiL$ and $\phiR$ by applying a linear transformation (see Supplementary Material, Section 5 for more details).
Simulations reproduced the measurements displayed in Fig.~\ref{fig2}b--d to a good degree. Calculated critical currents were between $10$ and $270~\mathrm{nA}$, with diode efficiencies up to $\eta^\mathrm{max}\approx 25 \%$ and patterns in the 2D phase space closely resembling the experimental data. By comparing the simulation of $\eta$ in Fig.~\ref{fig6}d to the result previously obtained with the minimal model (Fig.~\ref{fig5}d), we note a reduction of $|\eta|$ and broadening of the features located near $(\phiL,\phiR)=(\pi,\pi)$, modulo $2 \pi$ (see black arrows in both figures). This effect, also clearly visible in Fig.~\ref{fig2}d, is mainly related to the sinusoidal components to the CPR of Eq.~\ref{eq:CPR:2}, where $T_\mathrm{M}$ is substantially smaller than $T_\mathrm{L,R}$, and marginally related to $\tau_\mathrm{M}$, only slightly smaller than $\tau_\mathrm{L,R}$. This is expected in the device under study, as the larger length of the S--M JJ (lithographically of $120~\mathrm{nm}$) compared to S--L and S--R ($50~\mathrm{nm}$) reduced both the transmission of the most transmissive modes and the number of channels with low transmission.

The switch-OFF configuration (Fig.~\ref{fig2}e--h) is investigated by further adjusting the numerical model based on the following considerations. When the right superconducting loop is interrupted, the current flowing from terminal S to R does not have a direct path to D, but it must flow across the R--M junction. Thus, we introduce this junction in the model, with CPR $I_\mathrm{RM}$ also assumed to have the form of Eq.~\ref{eq:CPR:2}. Here the phase difference $\phiR - \phiM = \phiR$ is used instead of $\phi_\mathrm{S} - \phi_\alpha$ and parameters $\tau_\mathrm{RM}$ and $T_\mathrm{RM}$ substitute $\tau_\alpha$ and $T_\alpha$ (see schematic of Fig.~\ref{fig6}e). For these parameters we choose the values $\tau_\mathrm{RM}=0.97$ and $T_\mathrm{RM}=3.2$. The other consequence of interrupting the right loop is that $\phiR$ is not controlled externally with a magnetic flux, hence it is first calculated imposing the condition $I_\mathrm{SR}(\phiS-\phiR)=I_\mathrm{RM}(\phiR)$ (i.e., the current flowing from S to R equals that flowing from R to M), for any value of $\phiS$. Once $\phiR$ is determined, $\Icpn(\phiL)$ is computed using Eqs.~\ref{eq:Itot}, \ref{eq:Icpn} and \ref{eq:CPR:2}, leading to the result shown in Figs.~\ref{fig6}f and \ref{fig6}g. In agreement with Figs.~\ref{fig2}f and \ref{fig2}g, the model produces oscillations of $\Icpn$ as a function of $\phiL$ between $50~\mathrm{nA}$ and $240~\mathrm{nA}$, with a phase shift when reversing the current direction. This results in a diode efficiency (see Fig.~\ref{fig6}h) up to $40\%$, comparable to the measured value of $\approx 34\%$ and exhibiting an oscillating behavior depending on $\phiL$, similar to that in Fig.~\ref{fig2}h. 

The higher $\eta$ obtained in the switch-OFF case is understood by considering the higher asymmetry in the supercurrent distribution obtained in this setting. In fact, the supercurrent flowed directly from S to the common node D only via S--L and S--M, which had largely different transmissions due to the device geometry, while it had to traverse both S--R and R--M to reach D via R. This realized a strongly asymmetric situation, where junctions with different harmonic components led to large diode efficiencies \cite{Souto2022}.
When the switch was ON (Figs.~\ref{fig2}d and \ref{fig6}d) and junction S--R directly connected S to D, the supercurrent distribution became more symmetric, and spatial-inversion symmetry was broken by the different phases tuned with the loops: that is, local magnetic fluxes broke both time-reversal and spatial-inversion symmetry. Similar arguments apply to the case of Figs.~\ref{fig4}b and \ref{fig4}d, however setting gates to negative values to deplete parts of the semiconducting region reduced the maximum switching current, which also resulted in a decrease of the JDE efficiency (see Fig.~\ref{fig3}). The absence of JDE for the situation of Fig.~\ref{fig4}f is explained by considering that, with terminal M blocked, the supercurrent flowed in S--L and S--R only, which were almost balanced channels. Furthermore, phase tuning could not break spatial-inversion symmetry with M blocked, effectively resulting in a single superconducting loop between L and R. This situation therefore realized the balanced SQUID device of Ref.~\cite{Souto2022}, showing no JDE despite the nonsinusoidal CPR of the individual junctions.

We finally note that our model captured the main experimental results well without invoking hybridized ABSs in the semiconducting region, even though such states were measured in the same devices by means of local tunneling spectroscopy \cite{Coraiola2023}. There, the terminal S was weakly coupled to the multiterminal region and ABSs arising in the L--M and M--R junctions overlapped. In the present case, these JJs were not required to describe the experimental data (except for the switch-OFF configuration, where the M--R junction was not phase-controllable), since the supercurrents flowing from S to L, M and R had direct superconducting paths to D, without traversing L--M or M--R; this also excluded the contribution of hybridized modes between the two JJs. Furthermore, supercurrents were transported by several ABSs present in our devices, most of which solely depended on phase differences between pairs of superconducting contacts. Presumably, there were very few hybridized ABSs and they were localized close to the center of the multiterminal region. Thus, in the geometry studied here, they did not provide a significant contribution to the total supercurrent.

\section*{Conclusions}
We reported switching current measurement of a 4TJJ in a InAs/Al heterostructure hosting ABSs with large transmission probabilities, resulting in nonsinusoidal CPRs between pairs of terminals. The switching current measured between two contacts showed a strong dependence on the bias current direction, resulting in a JDE. The JDE efficiency could be widely controlled---both in amplitude and sign---by magnetic fluxes, independently tuned via integrated flux-bias lines, and gate electrodes, which routed the supercurrent to different transport channels. No external magnetic fields were required. In a first gate setting, where transport through the entire semiconductor region was allowed, the JDE efficiency was periodically modulated by magnetic fluxes, with peak values reaching $\eta\approx\pm 21\%$, including large regions in parameter space with $\eta\approx0$. When a superconducting arm was interrupted, introducing a larger asymmetry in the supercurrent distribution, a peak efficiency $\eta\approx\pm 34\%$ was reached. The 4TJJ was mapped onto a simple bi-SQUID geometry, with three parallel JJs containing ABSs with large transmission probability. A theoretical model reproduced the experimental observations to a good degree, including switching current and diode efficiency patterns. In our devices, the JDE is a consequence of the nonsinusoidal CPR and the multiterminal geometry, which allows breaking of spatial-inversion symmetry by controlling the magnetic fluxes in the loops. Unlike realizations based on a single loop, an asymmetry between junctions is not required. Our work highlights the potential of phase-tunable multiterminal JJs to engineer JDE with large and widely controllable efficiencies, without the need for exotic materials or external magnetic fields, and underscores the role of these devices as a versatile platform for upcoming applications.

\section*{Methods}

\subsection*{Materials and Fabrication}

Devices were realized in a III--V heterostructure grown by molecular beam epitaxy on an InP (001) substrate \cite{Cheah2023}. The semiconducting stack (starting from the substrate) consisted of a 1100 nm thick step-graded InAlAs buffer layer, a 6 nm thick $\mathrm{In_{0.75}Ga_{0.25}As}$ layer, an 8 nm thick InAs layer, a 13 nm thick $\mathrm{In_{0.75}Ga_{0.25}As}$ layer and two monolayers of GaAs. On top, an epitaxial 15 nm thick Al layer was deposited in situ without breaking vacuum.
A two-dimensional electron gas (2DEG) was confined in the InAs and its properties were characterized via measurements performed in a Hall bar geometry, which gave an electron peak mobility of $18000~\mathrm{cm^{2}V^{-1}s^{-1}}$ at an electron sheet density of ${8 \times 10^{11}~\mathrm{cm^{-2}}}$. This resulted in an electron mean free path $l_{e}\gtrsim260~\mathrm{nm}$ and a superconducting coherence length of the 2DEG proximitized by the Al sheet of {$\xi_\mathrm{InAs} = \sqrt{\hbar v_\mathrm{F} l_{e}/\left(\pi \Deltait\right)} \sim 600~\mathrm{nm}$}, with $v_\mathrm{F}$ the Fermi velocity in the 2DEG and $\Deltait$ the induced superconducting gap.

In the fabrication process, large mesa structures were first isolated, suppressing parallel conduction between devices and across the middle regions of the superconducting loops. This was done by selectively etching the Al layer with Transene type D, followed by a second chemical etch to a depth of $\sim 380 ~ \mathrm{nm}$ into the III--V material stack, using a $220:55:3:3$ solution of $\mathrm{H_{2}O:C_{6}H_{8}O_{7}:H_{3}PO_{4}:H_{2}O_{2}}$. Next, features were defined in the Al layer by wet etching with Transene type D at $50\mathrm{^{\circ}C}$ for $4~\mathrm{s}$.
The dielectric, uniformly deposited on the entire chip by atomic layer deposition, consisted of a $3~\mathrm{nm}$ thick layer of $\mathrm{Al_2 O_3}$ and a $15~\mathrm{nm}$ thick layer of $\mathrm{HfO_{2}}$. Gate electrodes and flux-bias lines were defined by evaporation and lift-off. In a first step, $5~\mathrm{nm}$ of Ti and $20~\mathrm{nm}$ of Au were deposited to realize the fine features of the gates; in a second step, a stack of Ti/Al/Ti/Au with thicknesses $5~\mathrm{nm}$, $340~\mathrm{nm}$, $5~\mathrm{nm}$ and $100~\mathrm{nm}$ was deposited to connect the mesa structure to the bonding pads and to define the flux-bias lines.

\subsection*{Measurements Techniques}

Experiments were performed in a dilution refrigerator with a base temperature at the mixing chamber below $10~\mathrm{mK}$. 
The sample was mounted on a QDevil QBoard sample holder system, without employing any light-tight enclosure. Electrical contacts to the devices, excepts for the flux-bias lines, were provided via a resistive loom with QDevil RF and RC low-pass filters at the mixing chamber stage, and RC low-pass filters integrated on the QBoard sample holder. Currents were passed through the flux-bias lines via a superconducting loom with only QDevil RF filters at the mixing chamber stage. Signals were applied to all gates and flux-bias lines via home-made RC filters at room temperature.

In all electrical measurements, a bias current $\Isd$ was driven between terminal S and node D of the device by applying equal and opposite voltages to S and D via bias resistors, whose resistance was much larger than that of the device under study. The voltage drop across S and D was detected in a four-terminal configuration. Measurements of the differential resistance were performed with lock-in-amplifier techniques, by applying a fixed AC current $\delta I = 2.5~\mathrm{nA}$ to D in addition to the DC bias $\Isd$ and detecting the AC voltage $\delta V$ between S and D, thus obtaining the differential resistance $R \equiv \delta V / \delta I$. Measurements of $\Iswpn$ were done by periodically ramping $\Isd$ from $0$ to an amplitude $A$, where $A$ was positive or negative depending on whether $\Iswp$ or $\Iswn$ was measured; the absolute value of $A$ was adjusted depending on the gate configuration to be slightly larger than $\Iswpn$. The signal form was a sawtooth wave, applied at a frequency $f = 133~\mathrm{Hz}$ using a waveform generator. The voltage drop across S and D was measured with an oscilloscope (averaging 32 measurements), which detected the time interval $\Delta t$ where the voltage was below a threshold, hence allowing for the calculation of the switching current as $\Iswpn = |A| f \Delta t$.

The dilution refrigerator was equipped with a superconducting vector magnet which, despite not being utilized for the experiments, produced a small magnetic field offset. Therefore, small offsets in the flux-bias line currents $\IL$ and $\IR$ (up to $\sim 100~\mathrm{\mu A}$) were considered in datasets, in such a manner that the point where $\IL = \IR = 0$ corresponded to a point of the phase space where $\eta = 0$ and $\Iswpn$ were maximal, as expected when no magnetic fluxes thread the superconducting loops.

\vspace{8mm}
\section*{Data availability}
Data presented in this work will be available on Zenodo. The data that support the findings of this study are available upon request from the corresponding author.

\section*{Acknowledgments}
We acknowledge fruitful discussions with R.~Seoane Souto and H.~Weisbrich.
We thank the Cleanroom Operations Team of the Binnig and Rohrer Nanotechnology Center (BRNC) for their help and support.
W.W.~acknowledges support from the Swiss National Science Foundation (grant number 200020\_207538).
J.C.C.~acknowledges support from the Spanish Ministry of Science and Innovation (Grant No.\ PID2020-114880GB-I00) and thanks the Deutsche Forschungsgemeinschaft (DFG; German Research Foundation) via SFB 1432 for sponsoring his stay at the University of Konstanz as a Mercator Fellow.
W.B.~acknowledges support from the DFG via SFB 1432 (ID 425217212) and BE 3803/14-1 (ID 467596333).
F.N.~acknowledges support from the European Research Council (grant number 804273) and the Swiss National Science Foundation (grant number 200021\_201082).

\bibliography{Full_Bibliography}

\begin{thebibliography}{52}
\bibitem [{Not()}]{Note_Lgeom}%
  \BibitemOpen
  \href@noop {} {}\bibinfo {note} {Geometric inductance $L_\mathrm{geom}$ of a
  loop was calculated analytically using the following expression, valid for a
  rectangular loop: $L_\mathrm{geom} = \frac{\mu_0}{\pi} \Bigr[ -2(W+H) +2
  \sqrt{H^2 + W^2} - H \ln \left( \frac{H + \sqrt{H^2+W^2}}{W} \right) - W \ln
  \left( \frac{W + \sqrt{H^2+W^2}}{H} \right) + H \ln \left( \frac{4 H}{d}
  \right) + W \ln \left( \frac{4 W}{d} \right) \Bigr]$. Here, $H=26~\mathrm{\mu
  m}$ and $W=4.5~\mathrm{\mu m}$ are the dimensions of the loop,
  $d=1~\mathrm{\mu m}$ is the width of the Al strip forming the loop and
  $\mu_0$ is the vacuum magnetic permeability.}\BibitemShut {Stop}%
\end{thebibliography}

\clearpage
\newpage
\onecolumngrid

\newcounter{myc}
\renewcommand{\thefigure}{S.\arabic{myc}}

\section*{Supplementary Material}

\section{Additional Phase-Space Linecuts}

The switching current measured for positive bias current $\Iswp$ in the phase space (i.e., as a function of both flux-line currents $\IL$ and $\IR$), shown in Fig.~2b of the Main Text, is plotted again in Fig.~\ref{figS1}a over extended ranges of $\IL$ and $\IR$. Here, in addition to the phase-space linecut $\IR=100~\mathrm{\mu A}$ displayed in the Main Text (green marker), we consider the linecuts $\IR=20~\mathrm{\mu A}$ (white marker) and $\IR=-60~\mathrm{\mu A}$ (blue marker). Along these directions, we measured the differential resistance $R$ across the device as a function of $\IL$ and of the bias current $\Isd$ (always swept from $0$ to positive or negative values), and show the result in Figs.~\ref{figS1}b--d. The three linecuts revealed oscillations of the switching current as a function of $\IL$, with maximum switching currents of approximately $240~\mathrm{nA}$.
In all measurements, the supercurrent was nonreciprocal at positive and negative $\Isd$, confirming the presence of Josephson diode effect (JDE) with phase-dependent efficiency $\eta$, consistent with Fig.~2d of the Main Text.

\section{Results for Different $\Vl$, $\Vr$}

\begin{figure}[h]
	\includegraphics[width=0.5\columnwidth]{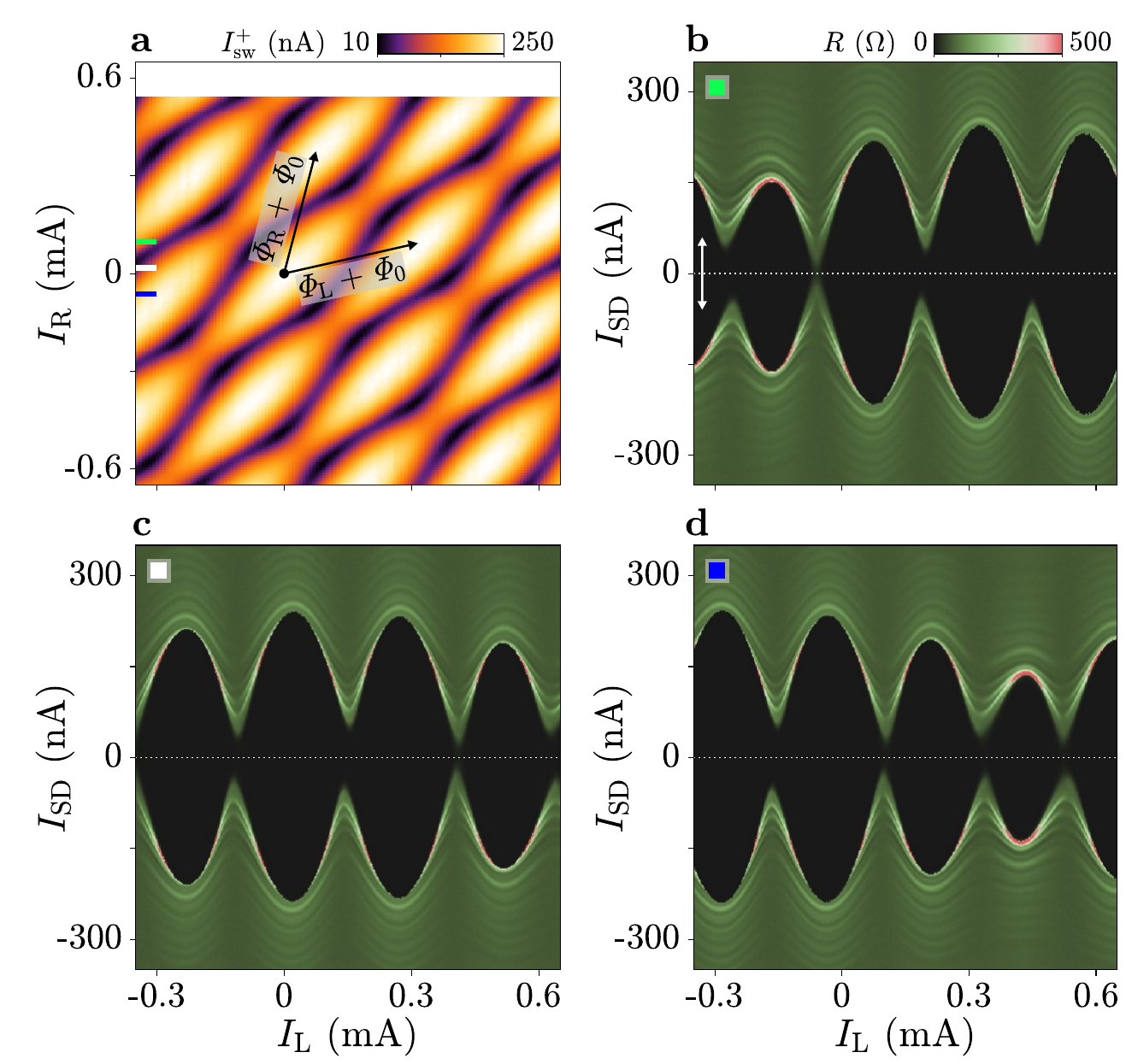}
	\caption{Phase-space linecuts at $\Vl=\Vr=-0.1~\mathrm{V}$. (a) Switching current $\Iswp$, measured for $\Isd>0$, as a function of flux-line currents $\IL$ and $\IR$, as in Fig.~2b of the Main Text. Colored markers indicate the position of $\IR=100~\mathrm{\mu A}$ (green), $\IR=20~\mathrm{\mu A}$ (white) and $\IR=-60~\mathrm{\mu A}$ (blue). (b--d) Differential resistance $R$ as a function of $\IL$ and $\Isd$ for $\IR=100~\mathrm{\mu A}$ (b), $\IR=20~\mathrm{\mu A}$ (c) and $\IR=-60~\mathrm{\mu A}$ (d), as indicated in (a). Each map is obtained by merging two datasets recorded with $\Isd$ ramping from $0$ to either positive or negative values [see white arrows in (b)].}
	\label{figS1}
\end{figure}

\begin{figure}[p]
	\includegraphics[width=0.5\columnwidth]{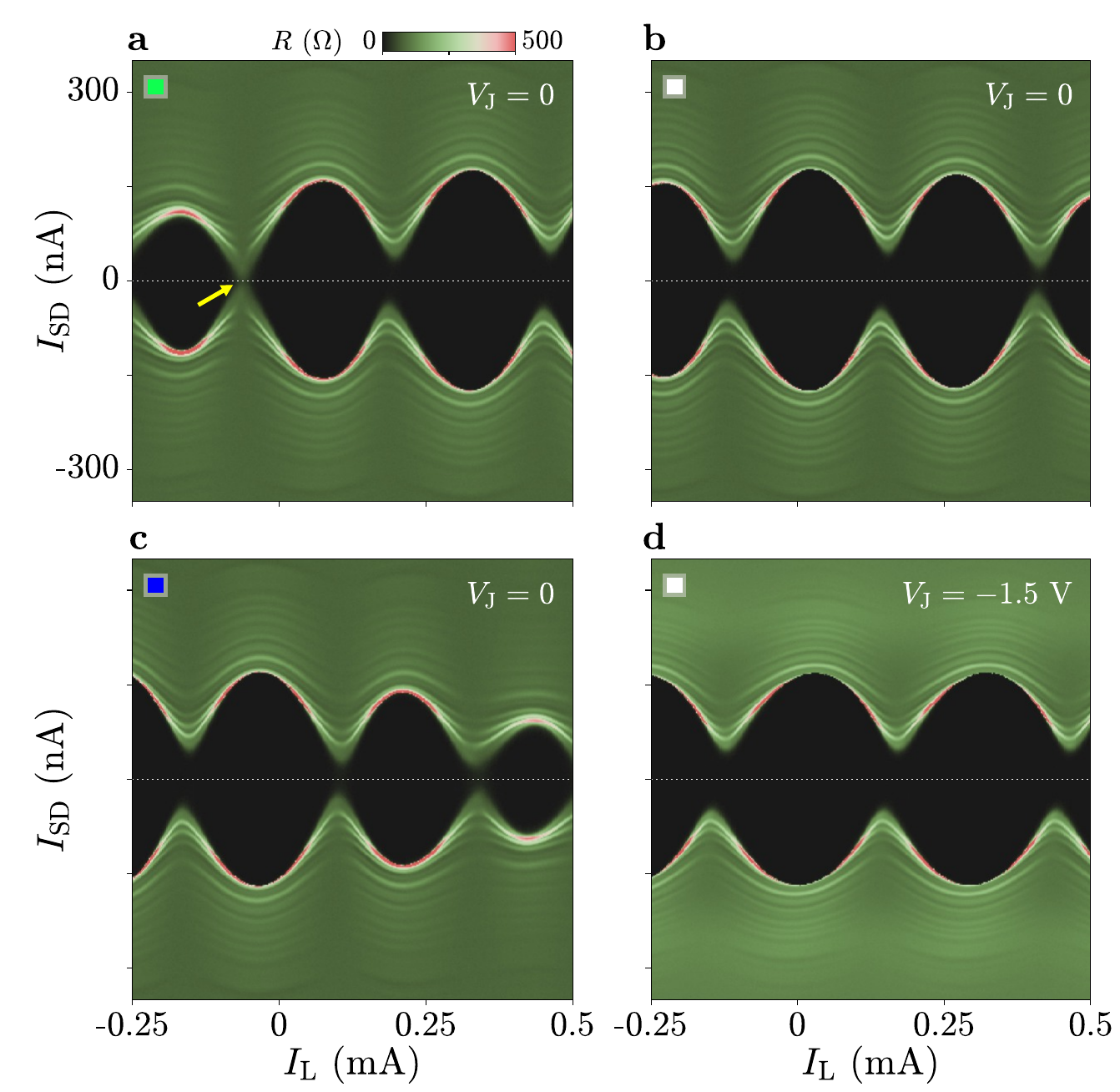}
	\caption{Phase-space linecuts at $\Vl=\Vr=-0.2~\mathrm{V}$. (a--c) Differential resistance $R$ as a function of $\IL$ and $\Isd$, for $\Vj=0$ (switch ON) and $\IR=100~\mathrm{\mu A}$ (a), $\IR=20~\mathrm{\mu A}$ (b) and $\IR=-60~\mathrm{\mu A}$ (c), as indicated in Fig.~\ref{figS1}a. In (a), a point where the switching current reaches zero is indicated by the yellow arrow. (d) As in (b), but for $\Vj=-1.5~\mathrm{V}$ (switch OFF).}
	\label{figS2}
\end{figure}

Figures ~\ref{figS2}a--c show three phase-space linecuts, indicated by the colored markers in Fig.~\ref{figS1}a, where the gate voltages $\Vl$ and $\Vr$ were set to $-0.2~\mathrm{V}$ (while $\Vl=\Vr=-0.1~\mathrm{V}$ in Fig.~2 of the Main Text and in Fig.~\ref{figS1}). The other gate voltages were kept to $\Vs=0.1~\mathrm{V}$, $\Vm=-0.15~\mathrm{V}$ and $\Vj=0$. The switching current oscillations qualitatively resembled those described for $\Vl=\Vr=-0.1~\mathrm{V}$ (Fig.~\ref{figS1}) but with reduced amplitude, as the maximum switching current was approximately $170~\mathrm{nA}$ in agreement with Fig.~3b of the Main Text. Regions of vanishing supercurrent (see for example the yellow arrow in Fig.~\ref{figS2}a) were more prominent in this gate configuration, consistent with the larger normal-state resistance.
The linecut of Fig. ~\ref{figS2}b ($\IR=20~\mathrm{\mu A}$) was also measured with the switch junction voltage set to $\Vj=-1.5~\mathrm{V}$ (switch OFF, Fig.~\ref{figS2}d), yielding a picture very similar to Fig.~2e of the Main Text. With respect to the case where $\Vl=\Vr=-0.1~\mathrm{V}$, we confirm a reduction of the maximum supercurrent (up to about $170~\mathrm{nA}$) and diode efficiency (up to approximately $25\%$).

\begin{figure}[p]
	\includegraphics[width=0.75\columnwidth]{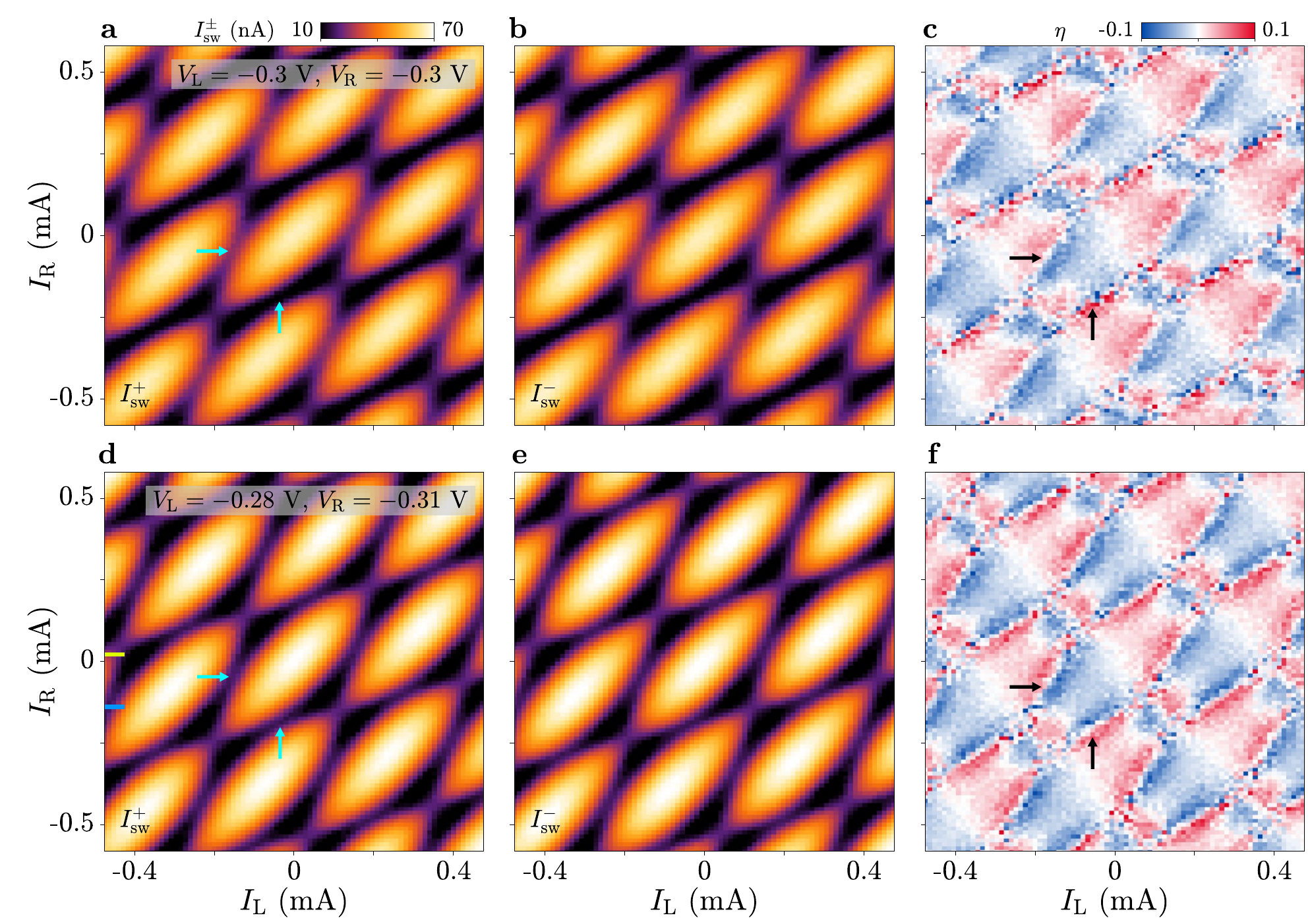}
	\caption{Comparing symmetric and asymmetric gate configurations. (a, b) Switching currents $\Iswp$ and $\Iswn$ (respectively) as functions of flux-line currents $\IL$ and $\IR$ for symmetric gate voltages $\Vl=\Vr=-0.3~\mathrm{V}$. The other gate voltages were kept to $\Vs=0.1~\mathrm{V}$, $\Vm=-0.15~\mathrm{V}$ and $\Vj=0$. The cyan arrows indicate the switching current minima described in the text. (c) Superconducting diode efficiency $\eta$ extracted from (a) and (b) as a function of $\IL$ and $\IR$. Regions indicated by the black arrows are discussed in the text. (d--f) As in (a--c), but for asymmetric gate voltages $\Vl=-0.28~\mathrm{V}$ and $\Vr=-0.31~\mathrm{V}$. In (d), yellow and blue markers indicate the position of $\IR=20~\mathrm{\mu A}$ and $\IR=-140~\mathrm{\mu A}$, respectively.}
	\label{figS3}
\end{figure}

The results shown thus far were acquired with symmetric gate voltages $\Vl=\Vr$, except in Fig.~4a--d of the Main Text where either gate was strongly depleted. In these configurations, the currents flowing into terminals L and R were relatively symmetric, as supported by our simulations (Fig.~6), where a good fit was found for $\tau_\mathrm{L}=\tau_\mathrm{R}$ and $T_\mathrm{L} \approx T_\mathrm{R}$. 
For $\Vl=\Vr=-0.3~\mathrm{V}$, we observed switching current patterns that were slightly asymmetric along the two periodicity directions, as shown in Figs.~\ref{figS3}a and \ref{figS3}b (see cyan arrows in panel a). This was likely due to different lever arm of the gates at voltages $\Vl$ and $\Vr$, which created an imbalance in the device for intermediate gate voltages. By setting $\Vl=-0.28~\mathrm{V}$ and $\Vr=-0.31~\mathrm{V}$, a more balanced situation was restored (see Figs.~\ref{figS3}d and \ref{figS3}e).
This effect was also visible in the superconducting diode efficiencies, shown in Figs.~\ref{figS3}c and \ref{figS3}f for the two cases: while the maximum $|\eta|$ slightly varied depending on the periodicity axis at $\Vl=\Vr=-0.3~\mathrm{V}$, it was substantially more symmetric in the two directions at $\Vl=-0.28~\mathrm{V}$, $\Vr=-0.31~\mathrm{V}$ (see black arrows in panels c and f, respectively).

\begin{figure}[p]
	\includegraphics[width=0.75\columnwidth]{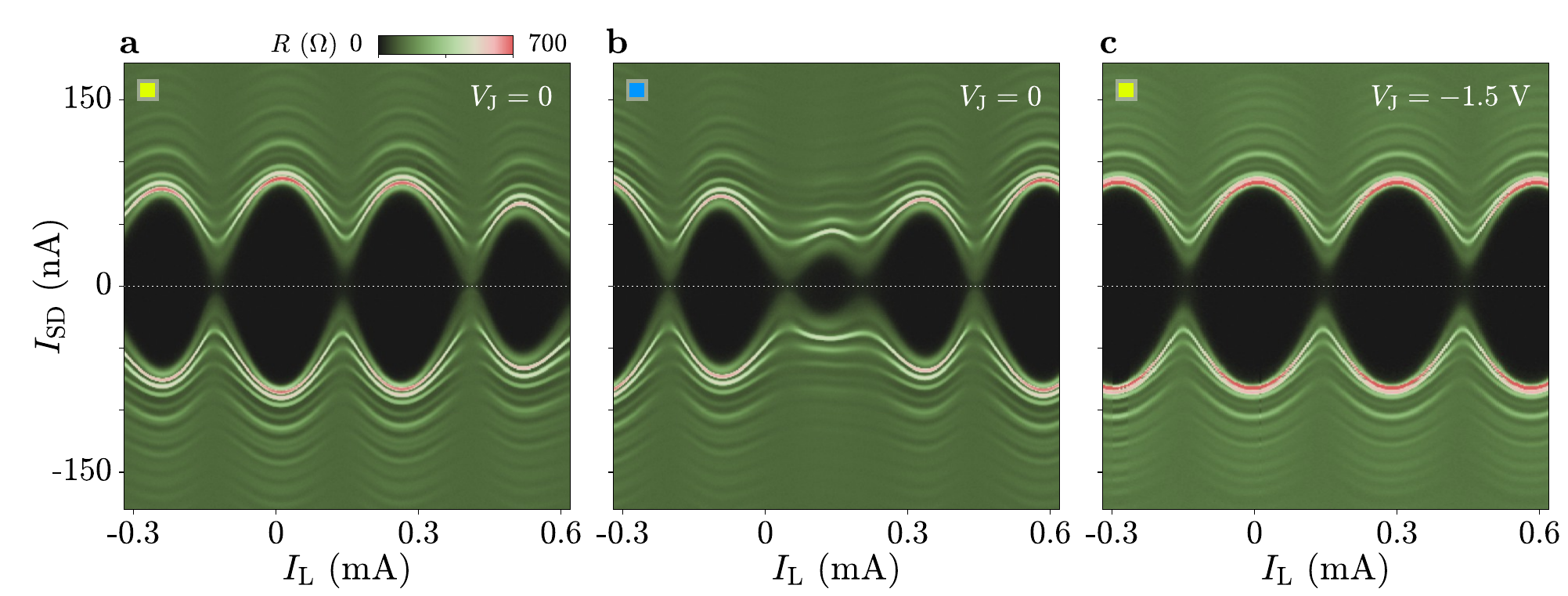}
	\caption{Phase-space linecuts at $\Vl=-0.28~\mathrm{V}$ and $\Vr=-0.31~\mathrm{V}$. (a) Differential resistance $R$ as a function of $\IL$ and $\Isd$ for $\Vj=0$ and $\IR=20~\mathrm{\mu A}$ (yellow marker in Fig.~\ref{figS3}d). (b) As in (a), but for $\IR=-140~\mathrm{\mu A}$ (blue marker in Fig.~\ref{figS3}d). (c) As in (a), but for $\Vj=-1.5~\mathrm{V}$ (switch OFF). Each map is obtained by merging two datasets recorded with $\Isd$ ramping from $0$ to either positive or negative values.}
	\label{figS4}
\end{figure}

Staying in the gate configuration with $\Vl=-0.28~\mathrm{V}$, $\Vr=-0.31~\mathrm{V}$, we defined two phase-space linecuts (colored markers in Fig.~\ref{figS3}d, corresponding to $\IR=20~\mathrm{\mu A}$ and $\IR=-140~\mathrm{\mu A}$) and measured $R$ along these directions as a function of $\IL$ and $\Isd$. The result is presented in Figs.~\ref{figS4}a and \ref{figS4}b, showing switching current oscillations with maxima of about $70~\mathrm{nA}$ and minima where the supercurrent vanished. Nonreciprocity was still present in the switching current depending on the sign of $\Isd$, despite less markedly than for the previous gate configurations (where $\Vl$ and $\Vr$ were set to less negative values), consistent with diode efficiencies up to approximately $10 \%$ reported in Fig.~\ref{figS3}f.
Finally, the gate voltage $\Vj$ was set to $-1.5~\mathrm{V}$ to operate with the switch OFF, and supercurrent oscillations were measured along the linecut at $\IR=20~\mathrm{\mu A}$ (see Fig.~\ref{figS4}c), also revealing maximum switching currents of close to $70~\mathrm{nA}$.

\section{Extracting the Diode Efficiency as a Function of $\Vl$, $\Vr$}

\begin{figure}[p]
	\includegraphics[width=\columnwidth]{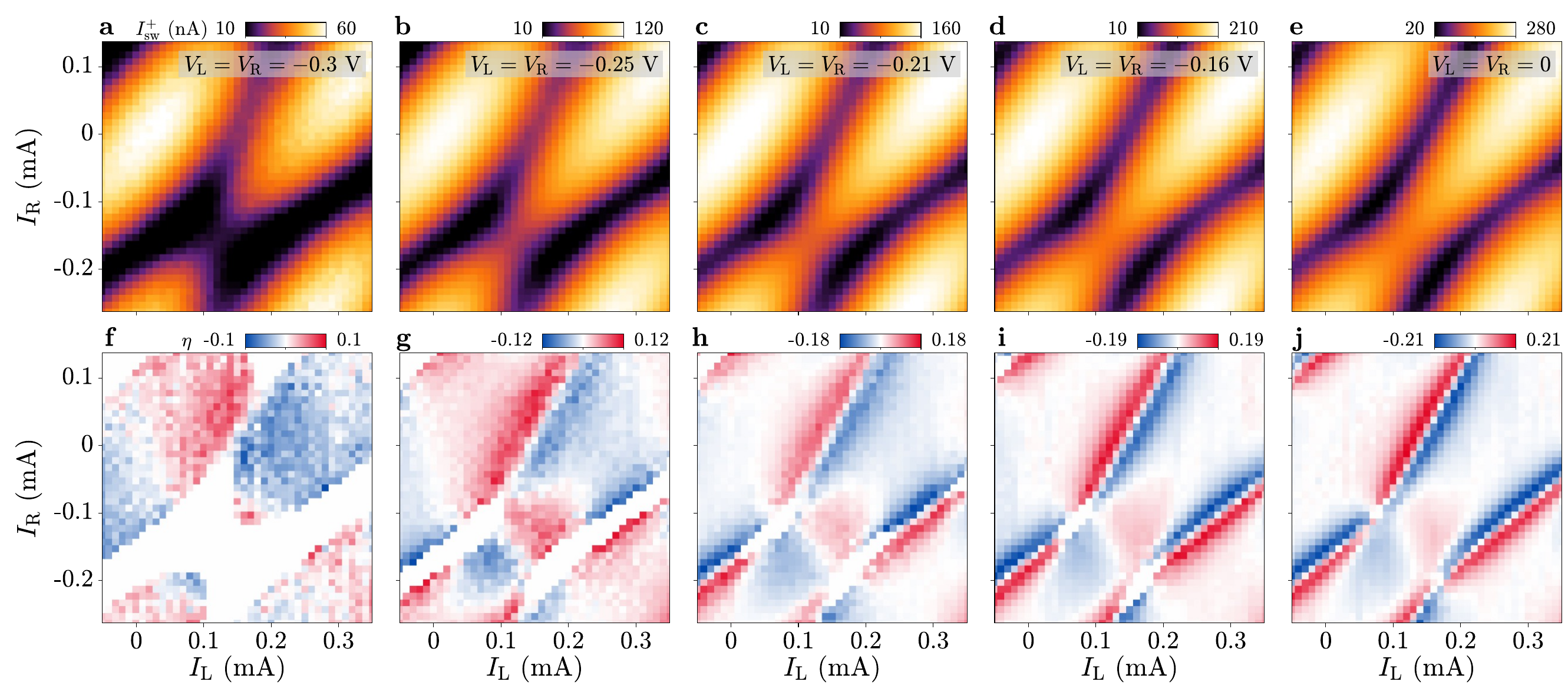}
	\caption{Results for varying $\Vl$, $\Vr$ and extraction of diode efficiencies. (a--e) Switching current $\Iswp$, measured for $\Isd>0$, as a function of flux-line currents $\IL$ and $\IR$ for five settings of $\Vl=\Vr$, indicated on the subfigures. (f--j) Diode efficiency extracted from (a--e) and the corresponding measurements of $\Iswn$ (not shown), respectively, as a function of $\IL$ and $\IR$. The other gate voltages were kept to $\Vs=0.1~\mathrm{V}$, $\Vm=-0.15~\mathrm{V}$ and $\Vj=0$.}
	\label{figS5}
\end{figure}

In Fig.~3b of the Main Text, we showed the gate-tunability of the JDE in our devices by presenting the dependence of the maximum diode efficiency $\eta^\mathrm{max}$ as a function of the gate voltages $\Vl=\Vr$. In the switch-ON configuration ($\Vj=0$), each data point was extracted from measurements of $\Iswp$ and $\Iswn$ taken as a function of both $\IL$ and $\IR$. Figures \ref{figS5}a--e and \ref{figS5}f--j display $\Iswp$ and $\eta$ obtained using Eq.~1 of the Main Text for five selected gate configurations ranging between $\Vl=\Vr=-0.3~\mathrm{V}$ and $\Vl=\Vr=0$. Since values close to the limit of detection ($\sim 10~\mathrm{nA}$) gave large variability of the extracted $\eta$ (see for example Figs.~\ref{figS3}c and \ref{figS3}f in the regions of phase space where $\Iswp$ was $\gtrsim 10~\mathrm{nA}$), regions where $\Iswp$ was lower than a threshold of $20~\mathrm{nA}$ were not considered. We note that the exact choice of the threshold did not significantly alter the result of the extraction. In each gate configuration, $I_\mathrm{sw}^\mathrm{max}$ and $\eta^\mathrm{max}$ were obtained by considering the 99.9th percentile of $\Iswp$ and $\eta$, respectively, throughout the corresponding maps.
In switch-OFF case ($\Vj=-1.5~\mathrm{V}$), included in Fig.~3b of the Main Text, the extraction was simplified for two reasons: first, since only one periodicity axis remained in the phase space, $\Iswpn$ could be measured along a single direction (e.g., as a function of $\IL$ for any fixed $\IR$), thus improving both speed and resolution of the measurement; second, $\Iswpn$ did not approach the limit of detection very closely, therefore a threshold was not required to select the data. Similar to the switch-ON case, $I_\mathrm{sw}^\mathrm{max}$ and $\eta^\mathrm{max}$ were obtained by considering the 99.9 percentile of $\Iswp$ and $\eta$ for each gate setting.

\section{Results for Device 2}

\begin{figure}[p]
	\includegraphics[width=0.3\columnwidth]{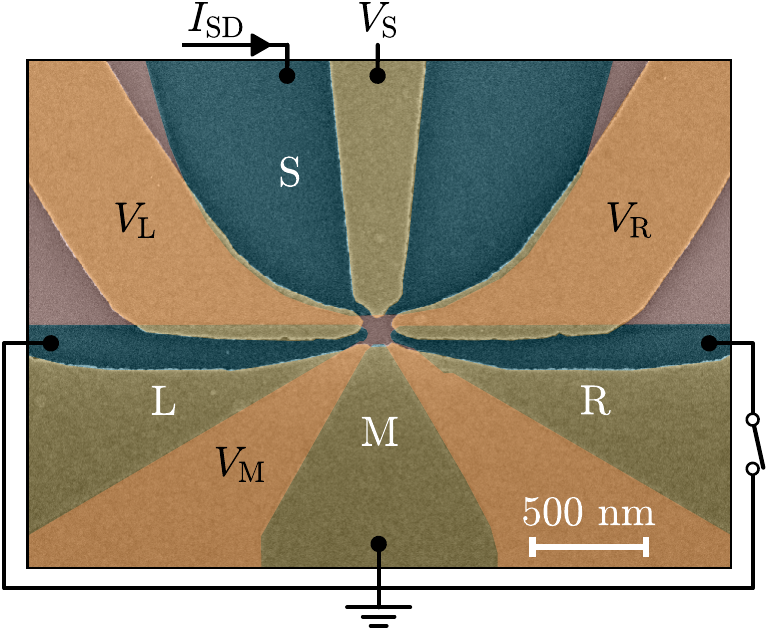}
	\caption{Device 2: false-colored scanning electron micrograph of the four-terminal Josephson-junction region. Color legend and labeled quantities are as in Figs.~1a and 1b of the Main Text. Circuit layout and loop geometry of Device 2 are lithographically identical to Device 1 (see Figs.~1a and 1b).}
	\label{figS6}
\end{figure}

To further study the JDE arising in a four-terminal Josephson junction (4TJJ) embedded in a double-loop geometry, we characterized a second device fabricated on the same chip. The circuit layout of Device 2 and the geometry of the two superconducting loops were lithographically identical to Device 1 (see Figs.~1a and 1c of the Main Text). The 4TJJ region, displayed in Fig.~\ref{figS6}, featured a different layout of terminal S (significantly wider than for Device 1), whereas terminals L, M and R were designed to be identical between the two devices. The shape of the gates was also varied, in particular for the gate energized by the voltage $\Vs$.

\begin{figure}[p]
	\includegraphics[width=0.75\columnwidth]{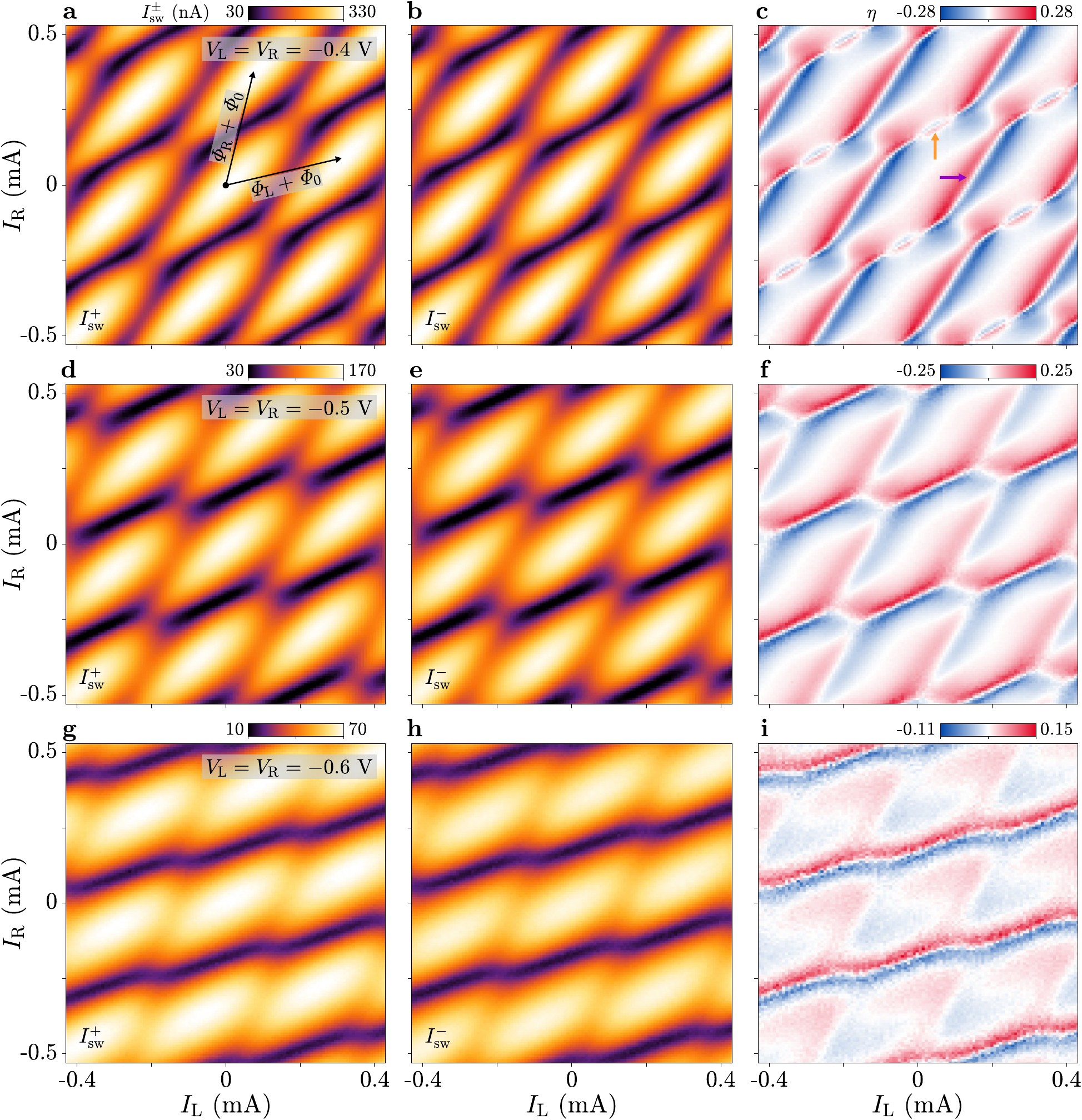}
	\caption{Phase- and gate-tunable Josephson diode effect in Device 2. (a, b) Switching currents $\Iswp$ and $\Iswn$, measured for $\Isd>0$ and $\Isd<0$ respectively, as functions of flux-line currents $\IL$ and $\IR$, for $\Vl=\Vr=-0.4~\mathrm{V}$. Directions of the black arrows in (a) indicate the periodicity axes, corresponding to the external magnetic fluxes $\PhiL$ and $\PhiR$ threading the two superconducting loops, while their length (one period) indicates the addition of one superconducting flux quantum $\Phio$ to the corresponding flux. (c) Superconducting diode efficiency $\eta$ obtained from (a) and (b) as a function of $\IL$ and $\IR$. Features indicated by the orange and purple arrows are discussed in the text. (d--f) As in (a--c), but for $\Vl=\Vr=-0.5~\mathrm{V}$. (g--i) As in (a--c), but for $\Vl=\Vr=-0.6~\mathrm{V}$. In all cases, the other gate voltages were set to $\Vs=0.15~\mathrm{V}$, $\Vm=-0.15~\mathrm{V}$ and $\Vj=0$.}
	\label{figS7}
\end{figure}

Current-bias measurements were performed as described for Device 1. Figure~\ref{figS7} shows the switching current $\Iswpn$ and the extracted diode efficiency $\eta$ as functions of the flux-line currents $\IL$ and $\IR$ for three configurations of the gate voltages $\Vl$ and $\Vr$ (set to a common value): $-0.4~\mathrm{V}$ (a--c), $-0.5~\mathrm{V}$ (d--f) and $-0.6~\mathrm{V}$ (g--i). The other gate voltages were $\Vs=0.15~\mathrm{V}$,~$\Vm=-0.15~\mathrm{V}$ and $\Vj=0$ (switch ON).
Switching current oscillations in the 2D phase space at $\Vl=\Vr=-0.4~\mathrm{V}$ qualitatively resemble those observed for Device 1 in Figs.~2b and 2c of the Main Text. The larger current maxima and minima, occurring despite voltages were applied in a more negative range than in Device 1, are understood by considering the different geometry of terminal S, that is compatible with a higher number of conduction channels in the S--L and S--R junctions, and a stronger screening of the electric field generated by the gates. Another feature of Device 2, that was particularly visible at $\Vl=\Vr=-0.5~\mathrm{V},~-0.6~\mathrm{V}$, was the large asymmetry between the two periodicity directions $\PhiL$ and $\PhiR$. This is attributed to a smaller supercurrent flowing from S to L than from S to R, likely due to a combination of a smaller number and transmission of the modes on the left side of the device, even for symmetrically applied gate voltages. We note that the case with $\Vl=\Vr=-0.6~\mathrm{V}$ is qualitatively similar to Figs.~4c and 4d of the Main Text, where only $\Vl$ was depleted. The asymmetry was reflected in the diode efficiency $\eta$: while oscillations of $\eta$ in the phase space had some resemblance with those of Device 1 (see Figs.~2d of the Main Text), and also reached large maxima up to $|\eta|\approx28\%$, they showed different amplitude and features depending on which flux was varied. A prominent example is visible in Fig.~\ref{figS7}c, where features of $\eta$ near the phase-space points $(\PhiL, \PhiR)=(0, \Phio/2)$ (modulo $\Phio$), marked by the orange arrow, have different shape and smaller $\eta$ than those near $(\PhiL, \PhiR)=(\Phio/2, 0)$ (purple arrow). Conversely, for $\Vl=\Vr=-0.5~\mathrm{V},~-0.6~\mathrm{V}$, $\eta$ was larger in proximity $(\PhiL, \PhiR)=(0, \Phio/2)$, consistent with the earlier suppression of $\PhiL$ dependence as the gate voltages were lowered.

\begin{figure}[p]
	\includegraphics[width=0.75\columnwidth]{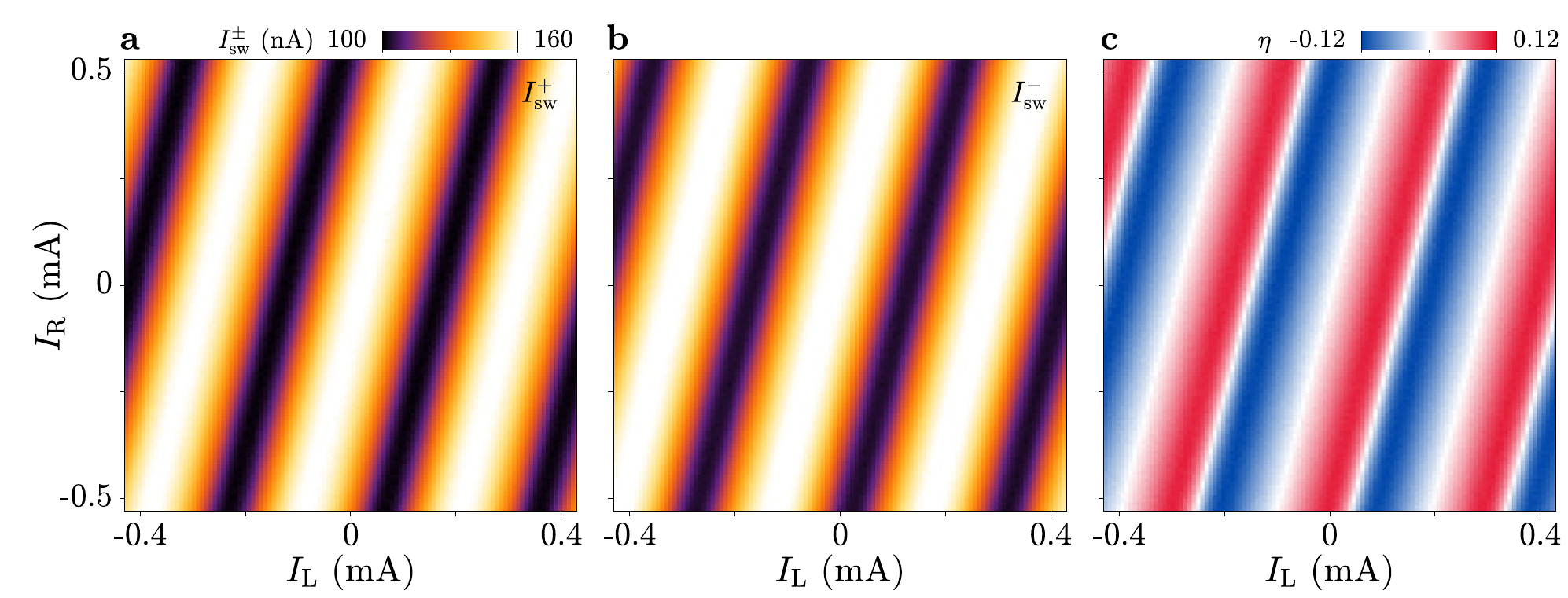}
	\caption{Josephson diode effect in Device 2 for $\Vj=-1.5~\mathrm{V}$ (switch OFF). (a, b) Switching currents $\Iswp$ and $\Iswn$, measured for $\Isd>0$ and $\Isd<0$ respectively, as functions of flux-line currents $\IL$ and $\IR$, for $\Vl=\Vr=-0.4~\mathrm{V}$. Here, $\Vl=\Vr=-0.5~\mathrm{V}$, $\Vs=0.15~\mathrm{V}$ and $\Vm=-0.15~\mathrm{V}$. (c) Superconducting diode efficiency $\eta$ obtained from (a) and (b) as a function of $\IL$ and $\IR$.}
	\label{figS8}
\end{figure}

Switching currents and diode efficiency of Device 2 were also measured in the switch-OFF configuration (\mbox{$\Vj=-1.5~\mathrm{V}$}), as shown in Fig.~\ref{figS8} for $\Vl=\Vr=-0.5~\mathrm{V}$. Oscillations of both $\Iswpn$ and $\eta$, suppressed along $\PhiR$, were similar to those reported for Device 1 in Figs.~2f--h of the Main Text. Interestingly, the maximum diode efficiencies of approximately $12\%$ were significantly lower than in the corresponding switch-ON case (Fig.~\ref{figS7}f), where $\eta^{\mathrm{max}} \approx 25\%$. Again, this is attributed to the asymmetric supercurrent distribution in Device 2, which led to small diode efficiencies along the $\PhiL$-axis and required tuning of $\PhiR$ to reach larger values in the case of Fig.~\ref{figS7}f . When the switch was OFF, control over $\PhiR$ was disabled, hence reducing $\eta^{\mathrm{max}}$.

The results shown for Device 2 were qualitatively captured by our extended theoretical model, with parameters $(\tau_\mathrm{L}, \tau_\mathrm{M}, \tau_\mathrm{R}, T_\mathrm{L}, T_\mathrm{M}, T_\mathrm{R})$ set to $(0.91, 0.87, 0.93, 3.5, 2, 6.5)$ for Figs.~\ref{figS9}a--c (Configuration 1), to $(0.44, 0.65, 0.87, 1, 1, 3.3)$ for Figs.~\ref{figS9}d--f (Configuration 2), and to $(0, 0.53, 0.67, 0.2, 0.2, 1)$ for Figs.~\ref{figS9}g--i (Configuration 3).
Furthermore, we simulated the device in the switch-OFF case using the same parameters as in Configuration 2 and, in addition, $\tau_\mathrm{RM}=0.75$, $T_\mathrm{RM}=4$; the result is displayed in Fig.~\ref{figS10}. In all cases, the phase axes $(\phiL,\phiR)$ were converted to flux-line-current axes $(\IL,\IR)$ using the transformation described in Section \ref{section_remapping}. We note that the chosen parameters were such that $\tau_\mathrm{L}<\tau_\mathrm{R}$ and $T_\mathrm{L}<T_\mathrm{R}$, supporting our interpretation of the asymmetry between $\PhiL$ and $\PhiR$ as a result of asymmetric supercurrent flow in the device between terminals L and R. In Configuration 3, the data was best described by using $\tau_\mathrm{L}=0$, suggesting that the high-transmission channel between S and L was depleted.
The measurements performed on Device 2 and the simulations done within the same model introduced for Device 1 support the generality of the observed phenomena, and in particular corroborate the presence and origin of the JDE in our device.

\section{Current-to-Flux Remapping}\label{section_remapping}

The b-SQUID geometry of our devices enables control over two superconducting phase differences (between terminals L and M, and between R and M), tuned by the currents in the two flux-bias lines, $\IL$ and $\IR$. These currents generated magnetic fluxes threading the two superconducting loops, $\PhiL$ and $\PhiR$, with a cross-coupling leading to an effect of $I_\mathrm{L(R)}$ on the flux through the opposite loop, $\mathit{\Phi}_\mathrm{R(L)}$. As a consequence, the $\PhiL$- and $\PhiR$-axes had a finite slope with respect to the $\IL$- and $\IR$-axes, as visible in Fig.~2b of the Main Text. We describe the cross-coupling by considering a mutual inductance matrix $\mathbf{M}$ that relates flux $\PhiL$ and $\PhiR$ to flux-line currents $\IL$ and $\IR$:
\begin{equation}\label{eqM}
	\begin{pmatrix}
		\PhiL \\
		\PhiR
	\end{pmatrix}
	= \mathbf{M} \cdot 
	\begin{pmatrix}
		\IL \\
		\IR
	\end{pmatrix}
	=
	\begin{pmatrix}
		M_\mathrm{LL}  & M_\mathrm{LR} \\
		M_\mathrm{RL}  & M_\mathrm{RR}
	\end{pmatrix}
	\cdot
	\begin{pmatrix}
		\IL \\
		\IR
	\end{pmatrix}.
\end{equation}
To quantify $\mathbf{M}$ (for Device 1), we consider the $\left( \PhiL, \PhiR \right)$ and $\left( \IL, \IR \right)$ coordinates of two points of the phase space [in addition to the origin, $\left( \PhiL, \PhiR \right) = \left( \IL, \IR \right) = \left( 0,0 \right)$], such as $\left( \Phio, 0 \right)$ and $\left( 0, \Phio \right)$, and substitute them in Eq.~\ref{eqM} \cite{Coraiola2023}. The resulting $4 \times 4$ equation system leads to the mutual inductance matrix:
\begin{equation}\label{eqM2}
	\mathbf{M} = 
	\begin{pmatrix}
		6.98 ~ \mathrm{pH}  & -1.40 ~ \mathrm{pH} \\
		-1.73 ~ \mathrm{pH} & 5.86 ~ \mathrm{pH}
	\end{pmatrix}
\end{equation}
for Device 1, and
\begin{equation*}
	\mathbf{M} = 
	\begin{pmatrix}
		7.03 ~ \mathrm{pH}  & -1.39 ~ \mathrm{pH} \\
		-1.67 ~ \mathrm{pH} & 5.80 ~ \mathrm{pH}
	\end{pmatrix}
\end{equation*}
for Device 2. We note that $\mathbf{M}$ is very similar between the two devices, as the loop and flux-line geometry was lithographically identical.
In Fig.~\ref{figS11}, we apply $\mathbf{M}$ from Eq.~\ref{eqM2} to perform a basis transformation and plot the data presented in Figs.~2b--d (shown again in Figs.~\ref{figS11}a--c) as a function of the magnetic fluxes $\PhiL$ and $\PhiR$ (see Fig.~\ref{figS11}d--f). 

\begin{figure}[p]
	\includegraphics[width=0.75\columnwidth]{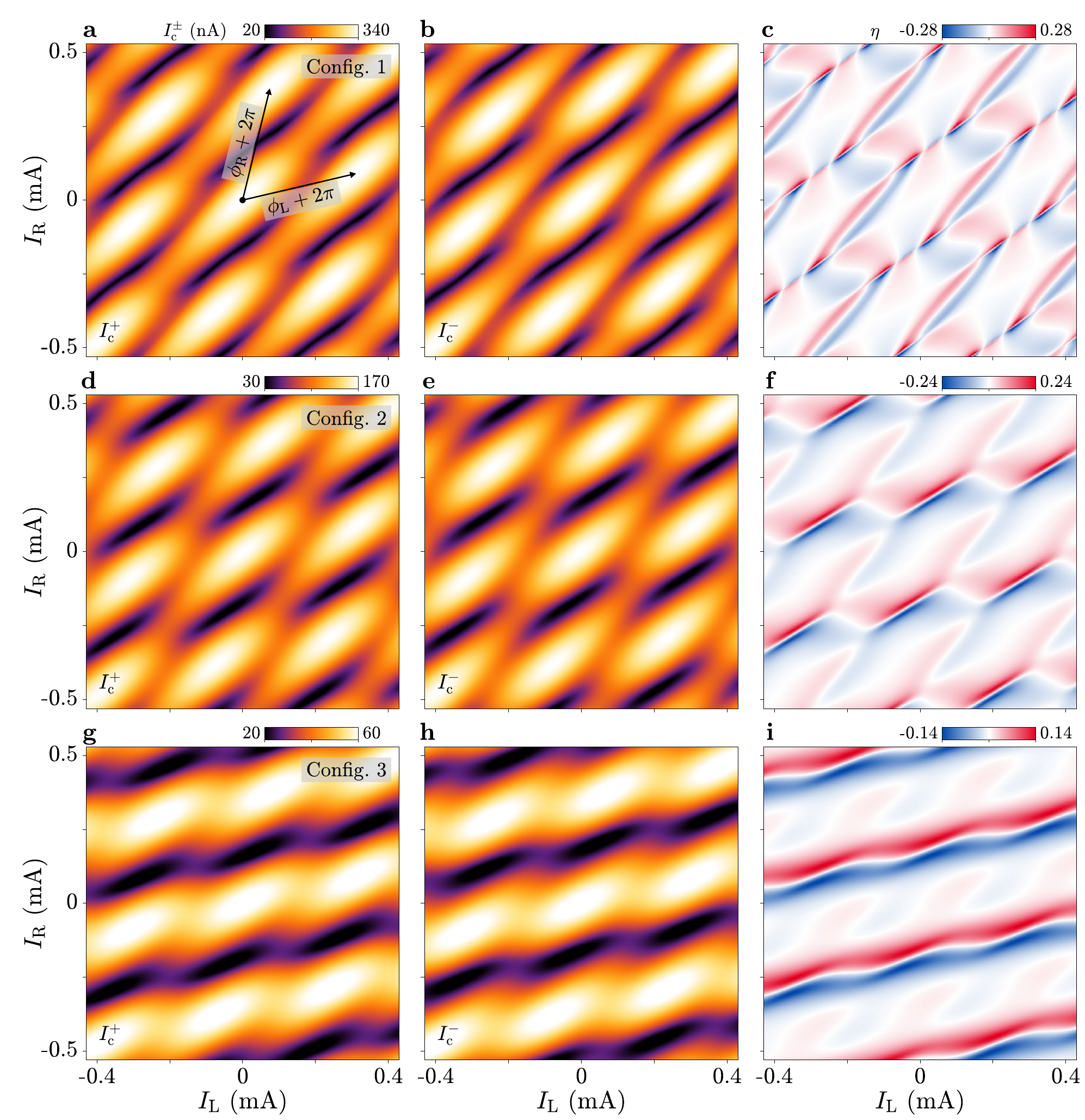}
	\caption{Simulations with the extended model for Device 2. (a, b) Critical currents $\Icp$ and $\Icn$, simulated for positive and negative current bias (respectively), as functions of flux-line currents $\IL$ and $\IR$, using the parameters of Configuration 1 (see text). The model is described in the Main Text. Currents $\IL$ and $\IR$ were obtained from the superconducting phase differences $\phiL$ (between terminals S and L) and $\phiR$ (between S and R) by applying a linear transformation (see Section \ref{section_remapping} for additional details). Phase axes $\phiL$ and $\phiR$, that are the periodicity directions, are indicated by the black arrows, whose length represents winding of the corresponding phase by $2 \pi$. (c) Diode efficiency $\eta$ extracted from (a) and (b) as a function of $\IL$ and $\IR$. (d--f) As in (a--c), but using the parameters of Configuration 2. (g--i) As in (a--c), but using the parameters of Configuration 3.}
	\label{figS9}
\end{figure}

\begin{figure}[p]
	\includegraphics[width=0.75\columnwidth]{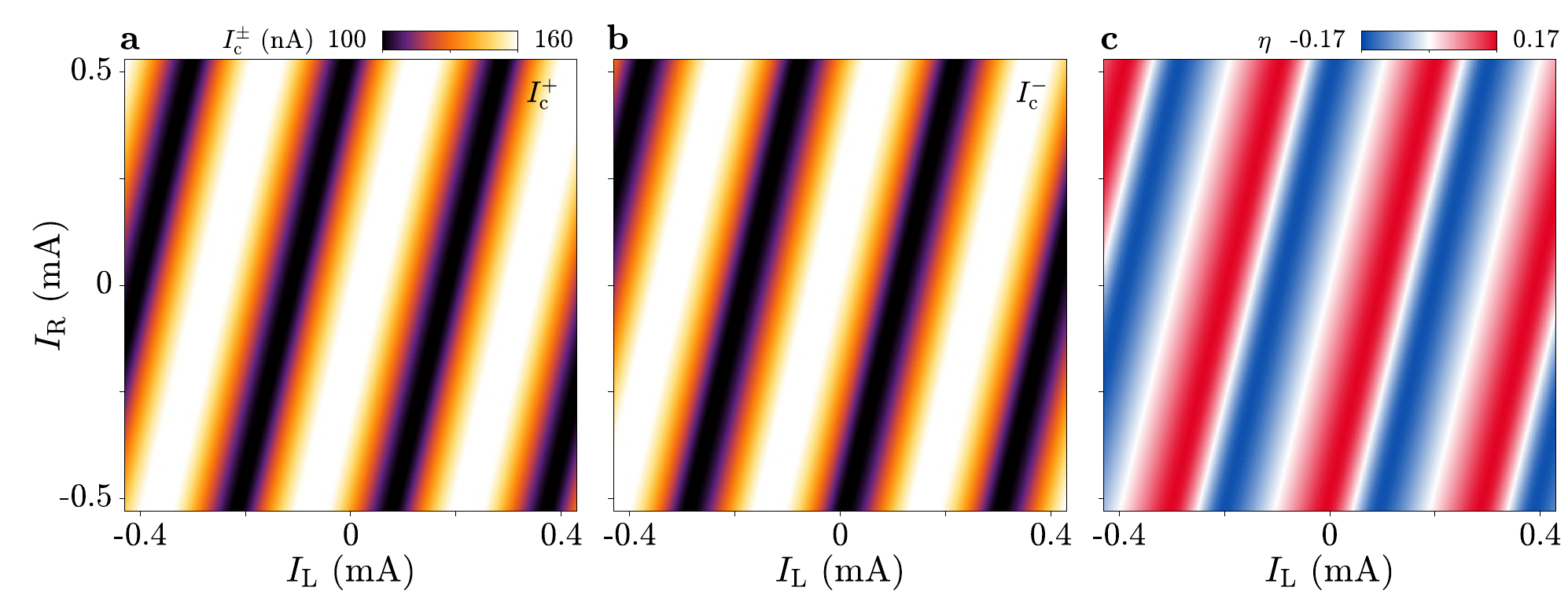}
	\caption{Simulations with the extended model for Device 2 in the switch-OFF case. (a, b) Critical currents $\Icp$ and $\Icn$, simulated for positive and negative current bias (respectively), as functions of flux-line currents $\IL$ and $\IR$, using the parameters of Configuration 2 (see text) and additional parameters $\tau_\mathrm{RM}=0.75$ and $T_\mathrm{RM}=4$. The model is described in the Main Text. (c) Diode efficiency $\eta$ extracted from (a) and (b) as a function of $\IL$ and $\IR$.}
	\label{figS10}
\end{figure}

\begin{figure}[p]
	\includegraphics[width=0.75\columnwidth]{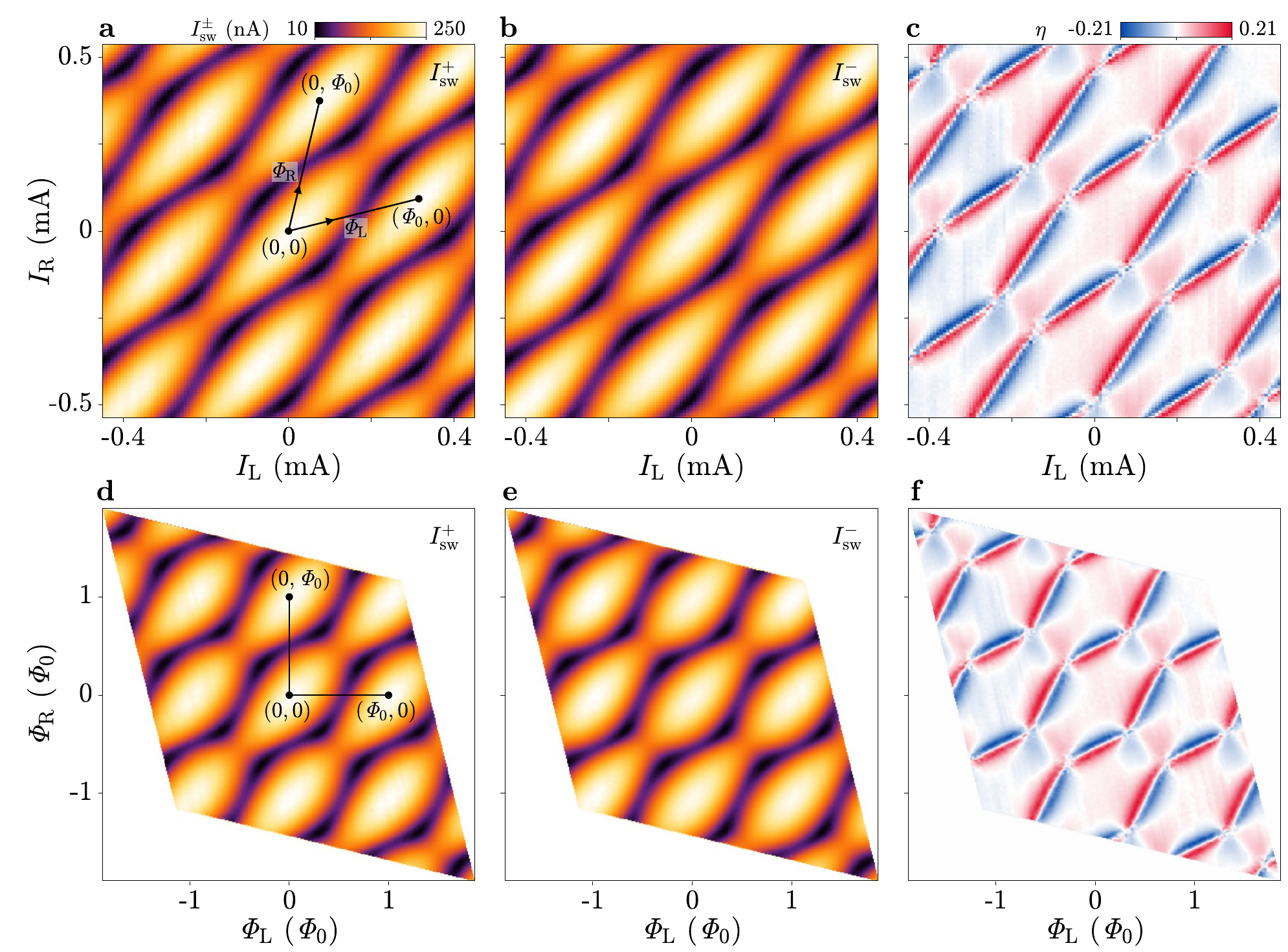}
	\caption{Linear transformation mapping the flux-line-current space into the flux space. (a--c) Switching currents $\Iswp$ and $\Iswn$ and diode efficiency $\eta$ as functions of flux-line currents $\IL$ and $\IR$, as in Figs.~2b--d of the Main Text. (d--f) Same datasets of (a--c) plotted as functions of the external magnetic fluxes $\PhiL$ and $\PhiR$, upon applying the linear transformation described in the text.}
	\label{figS11}
\end{figure}

The inverse transformation, obtained by inverting the mutual inductance matrix, was employed to map the phase space $(\phiL, \phiR)$ to the flux-line-current space $(\IL, \IR)$ for the simulated data of Figs.~6 of the Main Text, \ref{figS9} and \ref{figS10}. In this transformation, the phase differences were considered to be linearly related to the fluxes, thus $\mathit{\Phi}_\mathrm{L(R)} = \Phio \times \phi_\mathrm{L(R)} / 2 \pi$ were substituted in Eq.~\ref{eqM} to obtain a direct relation between phases and currents. This assumption is justified by noting that the inductance of the superconducting loops, comprising geometric and kinetic contributions $L_\mathrm{loop}=L_\mathrm{geom}+L_\mathrm{k}$, is negligible compared to the Josephson inductance $L_\mathrm{J}$ existing between any pairs of terminals in the 4TJJ. For each superconducting loop, the geometric inductance is estimated as $L_\mathrm{geom}\approx24~\mathrm{pH}$ \cite{Note_Lgeom}, while the kinetic inductance is calculated using the expression \cite{Annunziata2010}:
\begin{equation}
	L_\mathrm{k} = \frac{l}{w}\frac{h}{2 \pi^2}\frac{R_\square}{\mathit{\Delta}} \approx 100 ~ \mathrm{pH},
\end{equation}
where $l=57~\mathrm{\mu m}$ is the length of the loop perimeter, $w=1~\mathrm{\mu m}$ the width of the Al strip forming the loop, $h$ the Planck's constant, $R_\square\approx1.5~\Omega$ the normal state resistivity of the heterostructure stack (measured in Hall bar geometry where the Al was not removed) and $\Deltait \approx 180~\mathrm{\mu eV}$ is the superconducting gap of Al. Geometric and kinetic contributions lead to a total inductance $L_\mathrm{loop} \approx 124~\mathrm{pH}$. The Josephson inductance between two superconducting terminals (for example, L and M when considering the left loop) is estimated as $L_\mathrm{J}=\Phio/2 \pi I_\mathrm{c} \sim 3~\mathrm{nH}$, where $I_\mathrm{c} \sim 100~\mathrm{nA}$ gives the order of magnitude of the junction's critical current.

\setcounter{myc}{1}

\setcounter{myc}{2}

\setcounter{myc}{3}

\setcounter{myc}{4}

\setcounter{myc}{5}

\setcounter{myc}{6}

\setcounter{myc}{7}

\setcounter{myc}{8}

\setcounter{myc}{9}

\setcounter{myc}{10}

\setcounter{myc}{11}

\end{document}